1  **Characterization of a Thick Ozone Layer in Mars' Past**

2  J. Deighan[*], R.E. Johnson

3  Engineering Physics, 395 McCormick Road, University of Virginia, Charlottesville, VA, 22904

4  USA

5  Email Addresses: jid7v@virginia.edu[*],  rej@virginia.edu

6  Phone Number: 1-434-924-3244

**Abstract**

8      All three terrestrial planets with atmospheres support $O_3$ layers of some thickness. While

9 currently only that of Earth is substantial enough to be climatically significant, we hypothesize

10 that ancient Mars may also have supported a thick $O_3$ layer during volcanically quiescent periods

11 when the atmosphere was oxidizing. To characterize such an $O_3$ layer and determine the

12 significance of its feedback on the Martian climate, we apply a 1D line-by-line radiative-

13 convective model under clear-sky conditions coupled to a simple photochemical model. The

14 parameter space of atmospheric pressure, insolation, and $O_2$ mixing fraction are explored to find

15 conditions favorable to $O_3$ formation. We find that a substantial $O_3$ layer is most likely for

16 surface pressures of 0.3−1.0 bar, and could produce an $O_3$ column comparable to that of modern

17 Earth for $O_2$ mixing fractions approaching 1%. However, even for thinner $O_3$ layers, significant

18 UV shielding of the surface occurs along with feedback on both the energy budget and

19 photochemistry of the atmosphere. In particular, $CO_2$ condensation in the middle atmosphere is

20 inhibited and the characteristics of $H_2O$ dissociation are modified, shifting from a direct

21 photolysis dominated state similar to modern Mars to a more Earth-like state controlled by $O(^1D)$

22 attack.

## 1. Introduction

It is now known that all three atmosphere-bearing terrestrial planets—Venus, Earth, and Mars—posses an $O_3$ layer (Montmessin et al. 2011). In the $O_2$ rich atmosphere of modern Earth, $O_3$ has a significant impact on the solar and thermal radiation budgets, being the second most important absorber of shortwave radiation and third most important gas affecting longwave radiative forcing (Kiehl & Trenberth 1997). This results in $O_3$ providing substantial photochemical feedback to the climate (Hauglustaine et al. 1994). Stratospheric $O_3$ also performs the vital role of shielding surface organisms from damaging solar UV radiation. In contrast, the $O_3$ layers observed on Venus and Mars are of scientific interest only as probes into atmospheric photochemistry, being too thin to have significant climate feedback and providing hardly any UV attenuation.

But this may not always have been the case. The terrestrial planets' atmospheres have undergone major evolution since their formation (Hunten 1993), and it is known that Earth's atmosphere was not always so favorable to the formation of an $O_3$ layer (Kasting & Donahue 1980). By the same token, it is possible that the other terrestrial planets may have been *more* favorable to the formation of a stable $O_3$ layer than they are at present. If this was the case, the resulting energy budget and photochemistry feedbacks could be important in understanding the planets' climate evolution and habitability. In this paper we explore this possibility for Mars.

In order to identify conditions in Mars' past which would have favored a thick $O_3$ layer it is important to understand the reactions governing the production and destruction of $O_3$. Summarizing the arguments of Hiscox & Lindner (1997), we begin by examining the production of $O_3$ via the reaction

48
$$O + O_2 + M \rightarrow O_3 + M$$

49  where M is some third body, mostly likely $O_2$ or $N_2$ for Earth and $CO_2$ for Venus and Mars. The

50  kinetics of this reaction have a negative dependence on temperature, making $O_3$ production faster

51  at lower temperatures. And it is immediately obvious that increasing the concentrations [O], [$O_2$],

52  or [M] will increase the production rate of $O_3$. In the atmospheres we consider, much of the O

53  and $O_2$ are derived from the photolysis of $CO_2$, so the presence of a denser Martian atmosphere

54  will amplify the concentrations of all three species. On the other hand, destruction of the

55  oxidizing species O, $O_2$, and $O_3$ in the Martian atmosphere is largely controlled by catalytic

56  reactions involving $HO_x$ (H, OH, $HO_2$) produced by $H_2O$ photolysis. This means that a drier

57  atmosphere will have less $HO_x$, increasing [O] and [$O_2$] and reducing the destruction rate of $O_3$. A

58  thick and cold atmosphere will also contribute to reducing $HO_x$ by shielding $H_2O$ from UV and

59  lowering its vapor pressure, respectively. From this analysis, Hiscox & Lindner (1997)

60  concluded that a dense, cold, and dry atmosphere would be most favorable to the production of a

61  thick $O_3$ layer on Mars.

62      These conditions are somewhat at odds with those considered in many previous

63  paleoclimate studies of Mars. One of the primary drivers in studying Mars' past is explaining the

64  chemical and morphological evidence for surface modification by persistent liquid water (Carr &

65  Head 2010). Climate modeling struggles to achieve surface temperatures above 273 K simply by

66  assuming a thicker $CO_2$ atmosphere (Kasting 1991), and efforts to explain this discrepancy often

67  rely on additional greenhouse gases like $CH_4$, $H_2S$, and $SO_2$ (Halevy et al. 2009; S.S. Johnson et

68  al. 2008; Kasting 1997; Squyres & Kasting 1994). These species require a reducing atmosphere

69  and major geochemical source, conditions which are most easily met by large scale volcanic

70  activity, and result in a picture of early Mars not unlike Archean Earth (Halevy et al. 2007; S.S.

71  Johnson et al. 2009; Tian et al. 2010). Under such a warm, wet, and reducing scenario, oxidizing

72  species would be rapidly destroyed both in the atmosphere and in aqueous solution at the surface

73  (Kasting et al. 1979; Kasting & Walker 1981; Kasting et al. 1985). What little $O_3$ could exist

74  under such conditions would have negligible feedback on the climate.

75       However, Martian volcanism has been episodic throughout the planet's history, with the

76  most recent large scale caldera activity ending ~100–200 Ma (Neukum et al. 2010; Robbins et al.

77  2011). During volcanically quiescent geologic periods (such as the current time), any climate-

78  influencing volcanic gases would become depleted and the atmosphere would relax to a colder,

79  drier state. This would allow for the accumulation of oxidizing species such as $O_2$ and $O_3$

80  (Kasting 1995). Our characterization of $O_3$ climate feedback processes is pertinent to these

81  volcanically quiescent periods of Mars' history.

82       The photochemistry relevant to a thick $CO_2$ atmosphere lacking volcanism has been

83  examined by several authors (Hiscox & Lindner 1997; Rosenqvist & Chassefière 1995; Segura et

84  al. 2007; Selsis et al. 2002; Zahnle et al. 2008). The ability of trace amounts of a strong UV

85  absorber ($SO_2$) to modify the ancient Martian atmosphere's temperature structure has also been

86  demonstrated (Yung et al. 1997). However, the feedback between a photochemical product such

87  as $O_3$ and the planetary thermal structure has received less attention. Hiscox & Lindner (1997)

88  made the case that a Martian $CO_2$ atmosphere thick enough to support liquid water at the surface

89  would also produce an $O_3$ layer substantial enough to shield the surface from UV radiation and

90  contribute to the greenhouse effect. However, their analysis was rough and somewhat qualitative,

91  lacking detailed photochemical calculations. The studies of Rosenqvist & Chassefière (1995),

92  Segura et al. (2007), and Zahnle et al. (2008) used a fixed temperature profile for their
93  photochemical calculations. Selsis et al. (2002) did examine feedback between $O_3$
94  photochemistry and the temperature profile, but this was not the focus of the study and a
95  converged solution was not achieved due to excessively long run times.

96       Here we take a complementary approach, solving for the equilibrium temperature profile
97  using a radiative-convective model coupled to a simplified photochemical model. This provides a
98  first order estimate of stratospheric $O_3$ concentration while ensuring timely convergence. We
99  explore the parameter space of atmospheric pressure, insolation, and $O_2$ mixing fraction to search
100 for conditions likely to support a substantial $O_3$ layer and then examine the significance and
101 robustness of its feedback on the climate.

102

103 **2. Model Description**

104 2.1 Radiative-Convective Model

105      Radiative-convective equilibrium is solved for using a 1D line-by-line (LBL) model.  The
106 decision to use LBL is motivated by recent concerns that standard 1D models of ancient Mars,
107 such as used by Pollack et al. (1987) and Kasting (1991), suffer from serious parameterization
108 issues (Halevy et al. 2009; L. Kaltenegger, pers. comm., 2009; Wordsworth 2010a). There has
109 also been recent work demonstrating that correlated-k parameterizations commonly used for
110 Earth result in radiative forcing by $H_2O$ and $O_3$ which can deviate significantly from those given
111 by LBL calculations (Forster et al. 2011; Tvorogov et al. 2005). While LBL calculation is more
112 fundamentally correct than other methods and is important for benchmarking the accuracy of
113 alternative techniques, it has been rejected for many applications in the past because it is

114  computationally slow (Liou 2002). However, given new optimization techniques and the steady

115  advance in computing power, such models are becoming more common in planetary science

116  (Corrêa et al. 2005; Letchworth & Benner 2007; Kuntz & Höpfner 1999; Quine & Drummond

117  2002; Wells 1999).

118      In our model the atmosphere is divided into layers at a vertical resolution of 2 km, with

119  the domain extending from the planet's surface up to the altitude where atmospheric pressure

120  approaches $10^{-6}$ bar (0.1 Pa). The low pressure limit is chosen to avoid the complex effects

121  encountered in the upper atmosphere. These include the onset of the ionosphere, violation of

122  local thermodynamic equilibrium, transition of the dominant mass transfer mechanism from eddy

123  diffusion to molecular diffusion at the homopause, and transition of the dominant heat transfer

124  mechanism from radiation to conduction at the mesopause (Bougher et al. 2008, López-Valverde

125  et al. 1998).

126      Radiation is divided into two spectral regions: a UV region from 119 nm to 400 nm at 1

127  nm resolution and a visible/infrared (VIR) region ranging from 25000 $cm^{-1}$ (400 nm) to 0 $cm^{-1}$

128  with a non-uniform grid. The resolution of the pre-computed VIR grid varies from $2^{-15}$–1 $cm^{-1}$ in

129  order to resolve individual radiative lines at one half-width or better, and is optimized for the

130  input parameters of a given run (i.e. surface pressure, species' concentrations). The solar

131  radiation source in the UV is taken from the ASTM E-490 reference spectrum and VIR radiation

132  uses the synthetic spectrum provided by the AER group (Clough et al. 2005), which is generated

133  from the Kurucz (1992) solar source function. Solar fluxes are scaled by 1/4th to account for

134  global and diurnal averaging. Absorption by $CO_2$, $H_2O$, and $O_3$ is considered in both spectral

135  regions and $O_2$ is included in the UV.

136     UV cross-sections are obtained from various authors via the MPI-Mainz-UV-VIS

137     Spectral Atlas of Gaseous Molecules and Sander et al. (2011). Temperature dependent $CO_2$ UV

138     cross-sections are used via linear interpolation of data at 295 K and 195 K, being held constant

139     below 195 K.  The temperature sensitive Schumann-Runge bands of $O_2$ in the UV are computed

140     via the method of Minschwaner et al. (1992) and binned at 1 nm.

141     VIR cross-sections are calculated line-by-line using the 2008 HITRAN database

142     (Rothmann et al. 2009). Pressure broadening and line shifting of $H_2O$ lines by $CO_2$ instead of air

143     is implemented using the results of Brown et al. (2007). This increases pressure-broadened half-

144     widths by a factor of 1.7 on average relative to terrestrial air. Based on this, pressure-broadened

145     half-widths of $O_3$ lines in air are multiplied by 1.7, as data on $CO_2$ pressure-broadening of $O_3$ is

146     lacking. A Voigt line shape profile is used out to 40 Doppler half-widths from the line center and

147     beyond that a van Vleck–Weisskopf profile is used. For $H_2O$ and $O_3$ a Lorentzian wing cut-off of

148     25 $cm^{-1}$ from the line center is used, while sub-Lorentzian $CO_2$ wings are treated by using

149     empirical $\chi$-factors with a cut-off of 500 $cm^{-1}$ (Halevy et al. 2009; Wordsworth et al. 2010a). The

150     $\chi$-factor of Perrin & Hartmann (1989) is used in the range 0–3000 $cm^{-1}$, that of Tonkov et al.

151     (1996) is used for 3000–6000 $cm^{-1}$, and that of Meadows & Crisp (1996) is used > 6000 $cm^{-1}$.

152     For $H_2O$ and $O_3$ only the most common isotopologues are used ($H^{16}OH$, $^{16}O^{16}O^{16}O$), while

153     for $CO_2$ the three most common isotopologues are considered ($^{16}O^{12}C^{16}O$, $^{16}O^{13}C^{16}O$, $^{16}O^{12}C^{18}O$).

154     Also, for each layer weak radiative lines are excluded from calculations based on the criterion of

155     optical depth $\tau < 0.01$ at the line peak when integrated across the overhead atmospheric column.

156     These measures are taken for computational economy, and including rarer species and weaker

157     lines has a negligible (< 1 K) effect on the temperature profile.

158    Collision induced absorption (CIA) by $CO_2$ is accounted for using the "GBB"

159    parameterization of Wordsworth et al. (2010a), which is based on the experimental work of

160    Baranov et al. (2004) and Gruszka & Borysow (1998). The Chappuis and Wulf continua of $O_3$ in

161    the VIR are sampled at 1 cm$^{-1}$ from the MT_CKD continuum model (Clough et al. 2005).

162    Rayleigh scattering cross-sections for $CO_2$ are calculated from its refractive index. In the UV the

163    refractive index is taken from Bideau-Mehu et al. (1973), while in the VIR the results of Old et

164    al. (1971) are used. In both regions the King correction factor is taken from Alms et al. (1975).

165    For radiative transfer the plane-parallel approximation is used and treatment follows a

166    two-stream method, applying the technique of Toon et al. (1989) for scattering of solar radiation.

167    The planet's surface is taken to be a Lambertian reflector with an albedo of 0.20 for solar

168    radiation and an emissivity of 1.0 for thermal radiation. A moist pseudoadiabatic lapse rate is

169    used in the troposphere, with the surface as an $H_2O$ source at a specified relative humidity (RH).

170    Above the surface the mixing fraction of $H_2O$ is given a simple "step" profile, being held

171    constant until the condensation altitude, at which point its vapor pressure curve is followed up to

172    the cold trap based on the temperature profile. Above this the mixing fraction is again assumed to

173    be constant. Unless otherwise specified, a relatively dry surface value of RH=20% is used for all

174    the results given here, comparable to modern Mars (Zahnle et al. 2008). In regions where the

175    $CO_2$ atmosphere is super-saturated the lapse rate is set so that the temperature profile matches its

176    vapor pressure. Clear sky conditions are assumed.

177    For calculation of the tropospheric lapse rate, it is necessary to know the heat capacity $c_p$

178    of the atmosphere, dominated in this case by $CO_2$. From statistical mechanics, the zero-pressure

179    heat capacities of a rigidly rotating linear molecule with internal vibrational modes are

$$\frac{c_p}{R} = 1 + \frac{c_v}{R} = 1 + \frac{3}{2} + \frac{2}{2} + \sum_j \left(\frac{\theta_j}{T}\right)^2 \frac{e^{\frac{-\theta_j}{T}}}{\left(1 - e^{\frac{-\theta_j}{T}}\right)^2} \tag{1}$$

180  where $R$ is the universal gas constant, $T$ is the gas temperature, and $\theta_j$ is the characteristic

181  temperature of the $j$th vibrational mode. For $CO_2$ at low planetary temperatures ($T < 373$ K) only

182  the low energy vibrational modes $v_1 = 1330$ cm$^{-1}$ ($\theta_1 = 1910$ K) and doubly degenerate $v_2 = 667$

183  cm$^{-1}$ ($\theta_2 = 960$ K) need to be considered. Within this temperature range, Eq. (1) produces zero-

184  pressure $c_p$ values in excellent agreement (< 0.1% difference) with the values generated by the

185  NIST REFPROP program v9.0 (Lemmon 2010), which uses an elaborate $CO_2$ equation of state

186  (EOS) from Span & Wagner (1996) (see Fig. 1). For $T > 140$ K it also compares well with the $c_p$

187  parameterization derived from the Venusian atmosphere by Lebonnois et al. (2010). There is a

188  greater discrepancy with the expression used by Kasting (1991), however a review of the

189  literature reveals that this formulation originated in a paper by Eastman (1929) and is intended

190  for application from 300–2500 K.

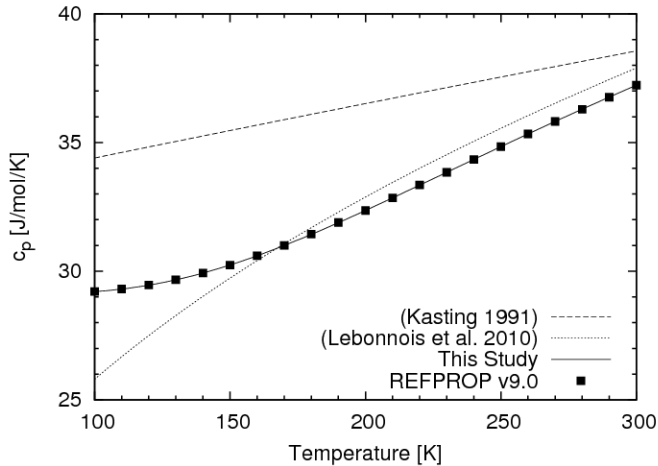

**Figure 1: Comparison of CO₂ specific heat capacity formulations.**

191

For non-zero pressures the heat capacity must be corrected for the non-ideal nature of the gas. In this we follow the appendix of Wordsworth et al. (2010b). To summarize, the Redlich-Kwong EOS of Li et al. (2006) is used, having the form

$$p = \frac{RT}{v-b} - \frac{a}{v(v-b)\sqrt{T}} \qquad (2)$$

where $p$ is pressure, $v$ is volume, and $a$ and $b$ are parameters specific to the gas under consideration. For the relatively low pressures considered here, this may be expressed as a Virial expansion

$$\frac{pv}{RT} \approx 1 + \frac{B}{v} = 1 + \frac{1}{v}\left(b - \frac{a}{RT^{3/2}}\right) \qquad (3)$$

which can be used to calculate the pressure correction to the heat capacity via the definition

$$\begin{aligned} c_p(p,T) &= c_p(0,T) - T\int_0^p \left(\frac{\partial^2 v}{\partial T^2}\right)_{p'} dp' \\ &= c_p(0,T) - T\,p\left(\frac{\partial^2 B}{\partial T^2}\right) \end{aligned} \qquad (4)$$

In the formulation of Li et al. (2006), the parameter $a$ is dependent on both $p$ and $T$, having the form $a = a_1(T) + a_2(T)p$. Because of the relatively low pressures relevant to Mars' past ($\leq$ 5 bar) we neglect the small pressure correction to $a$ and differentiate $B$ analytically. Calculating $c_p$ by this method agrees well with the NIST REFPROP program over the range of pressures and temperatures considered in this study.

Sensitivity testing with respect to the value of $c_p$ found that the surface temperature is not strongly affected by the use of our formulation as compared to that of Kasting (1991). Although

209 the lapse rate increases significantly, a concurrent lowering in the height of the tropopause limits

210 the surface temperature increase to at most a few K.

211       In our model the temperature profile is solved for by iteratively calculating the net flux

212 $F_{net}$ for each layer in the atmosphere and then setting the net flux to zero. This is done by

213 assuming that the absorbed flux is constant, and adjusting the emitted flux $F_{emt}$ by changing the

214 layer temperature $T$. To find the necessary step size $\Delta T$, an estimate of the derivative $dF_{emt}/dT$ is

215 required. We begin by taking the emitted flux to be

$$F_{emt} = \pi \int_0^\infty B(v,T)\left(1 - \mathrm{e}^{-\tau(v,T)}\right) dv \qquad (5)$$

216

217 By assuming that $\mathrm{e}^{-\tau}$ has a weak dependence on temperature over the interval $\Delta T$, we may

218 approximate the derivative as

$$\frac{dF_{emt}}{dT} \approx \pi \int_0^\infty \left[\frac{dB}{dT}\right]_v \left(1 - \mathrm{e}^{-\tau(v,T)}\right) dv \qquad (6)$$

219

220 For the conditions considered here, the radiatively active $v_2 = 667$ cm$^{-1}$ vibrational mode of $CO_2$

221 dominates thermal transfer where the atmosphere is in radiative equilibrium. This allows for two

222 more useful approximations. First, thermal emission occurs in the high energy tail of the Planck

223 function at the temperatures of interest ($T << \theta_2 = 960$ K), allowing the simplification

$$\frac{dF_{emt}}{dT} \approx \pi \int_0^\infty \frac{hv}{kT^2} B(v,T)\left(1 - \mathrm{e}^{-\tau(v,T)}\right) dv \qquad (7)$$

224

225 Second, by approximating the $v_2$ emission band as a $\delta$-function we may write

$$\frac{dF_{emt}}{dT} \approx \frac{hv_2}{kT^2} \pi \int_0^\infty B(v,T)\left(1 - \mathrm{e}^{-\tau(v,T)}\right) dv \qquad (8)$$

226

which is readily recognized as

$$\frac{dF_{emt}}{dT} \approx \frac{h\nu_2}{kT^2} F_{emt} \qquad (9)$$

228

This simple formulation is surprisingly accurate under the conditions for which it is derived, with typical errors < 1% compared to a numerical integration without approximations. After application of the derived $\Delta T$, a convective adjustment is applied in any layers where the lapse rate exceeds the pseudoadiabatic lapse rate.

An issue with convergence by this method arises when lower layers in the atmosphere become optically thick across the entire thermal IR. If there is a surplus of outgoing radiation at the top of the model, adjusting the surface temperature to achieve flux balance is no longer effective because changes must diffuse through the atmosphere up to space over several iterations. In the meantime, the surface receives little feedback and makes further changes, resulting in stability issues. This situation is typical of thick, warm, wet Martian atmospheres. Since modeling these climatic conditions is of general interest, the issue must be resolved. Our solution is to reduce temperatures throughout the optically thick region of the atmosphere along with the surface temperature. First we find the region between the surface and the altitude where the temperature drops below the planet's effective temperature $T_{eff}$. After the normal iteration technique to balance radiative flux and convection is applied, the layer in this region with the smallest $\Delta T$ value, $\Delta T_{min}$, is identified. The temperatures of all layers in the region are then adjusted

$$T_{adj} = T_{norm} - \Delta T_{min} + \Delta T_{surf} \qquad (10)$$

246

where $T_{adj}$ is the adjusted temperature, $T_{norm}$ is the normal unadjusted temperature, and $\Delta T_{surf}$ is the negative temperature change normally made at the surface. This technique ensures that the planet cools when there is a surplus of outgoing flux, but retains the relative temperature changes between atmospheric layers in each iteration.

251

2.2 Photochemical Model

The photochemical model is constructed as a series of atmospheric layers co-located with those of the radiative-convective model. The concentration profiles of $CO_2$, $CO$, $O_2$, and $H_2O$ are fixed, while $O_3$, $O$, and $O(^1D)$ are solved for by assuming local equilibrium. The reactions considered and their rate coefficients are listed in Table 1. Photolysis rates are derived from the UV/VIS flux profiles generated by the radiative-convective model, with quantum efficiencies taken from the review of Sander et al. (2011). The Kinetic-PreProcessor (KPP) code is used to solve for photochemical equilibrium (Daescu et al. 2003; Damien et al. 2002; Sandu et al. 2003).

For simplicity and clarity, eddy diffusion transport and $H_2O$ photochemistry after dissociation are neglected in the current model. While such a decision is not strictly accurate, it allows feedback mechanisms not associated with these processes to be examined independently and without obfuscation. The effects of neglecting these processes are most important in the troposphere and have qualitatively predictable consequences. These will be addressed later in the discussion.

For a real $CO_2$ atmosphere with no or little $H_2O$ photochemistry, i.e. a "dry" atmosphere, $CO$ and $O_2$ produced by photodissociation would exist at a stoichiometric ratio of 2:1 and could accumulate to mixing fractions of ~$10^{-2}$ due to the inefficiency of $CO_2$ recombination via $k_{10}$ in the absence of $HO_x$ species (McElroy & Donahue 1972; Nair et al. 1994; Parkinson & Hunten 1972). Under a "humid" atmosphere (modern Mars falls into this category) the $CO:O_2$ ratio declines and the mixing fractions can be reduced by orders of magnitude (on modern Mars the ratio is roughly 1:2 with mixing fractions on the order of $10^{-3}$). In addition, the planetary surface may act as a reductant source via volcanism (Kasting et al. 1979) and oxidant sink via weathering (Kasting & Walker 1981; Kasting et al. 1985; Zahnle et al. 2008). These act to increase CO concentration and decrease $O_2$ concentration. However, quantitatively evaluating the impact requires detailed analysis and is fraught with uncertainties about the climate, geochemistry, and volcanism of ancient Mars (Tian et al. 2010).

| Reactants | Products | Rate Coefficient | Reference |
|---|---|---|---|
| $CO_2 + hv \rightarrow CO + O$ | | $J_1$ | (Sander et al. 2011) |
| $CO_2 + hv \rightarrow CO + O(^1D)$ | | $J_2$ | (Sander et al. 2011) |
| $O_2 + hv \rightarrow O + O$ | | $J_3$ | (Sander et al. 2011) |
| $O_2 + hv \rightarrow O + O(^1D)$ | | $J_4$ | (Sander et al. 2011) |
| $O_3 + hv \rightarrow O_2 + O$ | | $J_5$ | (Sander et al. 2011) |
| $O_3 + hv \rightarrow O_2 + O(^1D)$ | | $J_6$ | (Sander et al. 2011) |
| $H_2O + hv \rightarrow H + OH$ | | $J_7$ | (Sander et al. 2011) |
| $O(^1D) + CO_2 \rightarrow CO_2 + O$ | | $k_8 = 1.63 \times 10^{-10} \ e^{60/T}$ | (Sander et al. 2011) |
| $O(^1D) + H_2O \rightarrow HO + HO$ | | $k_9 = 7.5 \times 10^{-11} \ e^{115/T}$ | (Sander et al. 2011) |
| $CO + O + M \rightarrow CO_2 + M$ | | $k_{10} = 2.2 \times 10^{-33} \ e^{-1780/T}$ | (Inn 1974) |
| $O + O + M \rightarrow O_2 + M$ | | $k_{11} = 2.5 \times 4.8 \times 10^{-33} \ (300/T)^{2\,*}$ | (Campbell & Gray 1973) |
| $O + O_2 + M \rightarrow O_3 + M$ | | $k_{12} = 2.5 \times 6.0 \times 10^{-34} \ (300/T)^{2.4\,*}$ | (Sander et al. 2011) |
| $O + O_3 \rightarrow O_2 + O_2$ | | $k_{13} = 8.0 \times 10^{-12} \ e^{-2060/T}$ | (Sander et al. 2011) |

**Table 1: Photochemical Reactions**

**\* Rate constant multiplied by 2.5 for increased efficiency of $CO_2$ as third body.**

279

280     For the purposes of this paper we simply assume that the $CO:O_2$ ratio is of order unity

281     and a range of prescribed mixing fractions from $10^{-5}$ to $10^{-2}$ is examined for these species. Based

282     on the work of Kasting & Donahue (1980), which quantified the impact of an $O_3$ layer on ancient

283     Earth, within this range $O_2$ should be stable enough against catalytic destruction by $HO_x$ to

284     maintain a roughly constant mixing fraction at all altitudes. This range of mixing fractions is also

285     consistent with the $CO_2$-$H_2O$ photochemical modeling results of Rosenqvist & Chassefière

286     (1995), and thus is appropriate for a low volcanic outgassing rate and slow surface weathering

287     under a cold, dry climate (Kasting 1995).

288

289     **3. Results & Discussion**

290     <u>3.1 Comparison with Previous Radiative-Convective Models</u>

291     As a validation of the radiative-convective model some test runs were performed without

292     photochemistry to compare with the results of selected models in the literature. Under modern

293     insolation a surface temperature of 273 K is achieved for a surface pressure of 2.5 bar, somewhat

294     higher than the 2.2 bar reported by Pollack et al. (1987). This may be explained by our inclusion

295     of $CO_2$ condensation (Kasting 1991) and differing treatment of CIA parameterization

296     (Wordsworth et al. 2010a). The differences in $H_2O$ profile structure and surface albedo between

297     the two models also have minor influences. To compare with the model of Wordsworth et al.

298     (2010a), the insolation is scaled by a factor of 0.75 at all wavelengths to approximate early solar

299     conditions and $H_2O$ is removed from the atmosphere. Under these conditions a peak surface

300     temperature of 215 K is found at a surface pressure of ~ 1 bar, and condensation of the 301  atmosphere at the surface occurs near a temperature of 208 K and pressure of just over 2.7 bar.

302  These results are comparable to those of Wordsworth et al. (2010a) for their "GBB" CIA

303  parameterization (Baranov et al. 2004; Gruszka & Borysow 1998) with sub-Lorentzian wings for

304  $CO_2$ lines. This is encouraging, as this model is the most similar to ours in the calculation of

305  radiative cross-sections.

306

307  3.2 Parameter Space Study Without Photochemistry

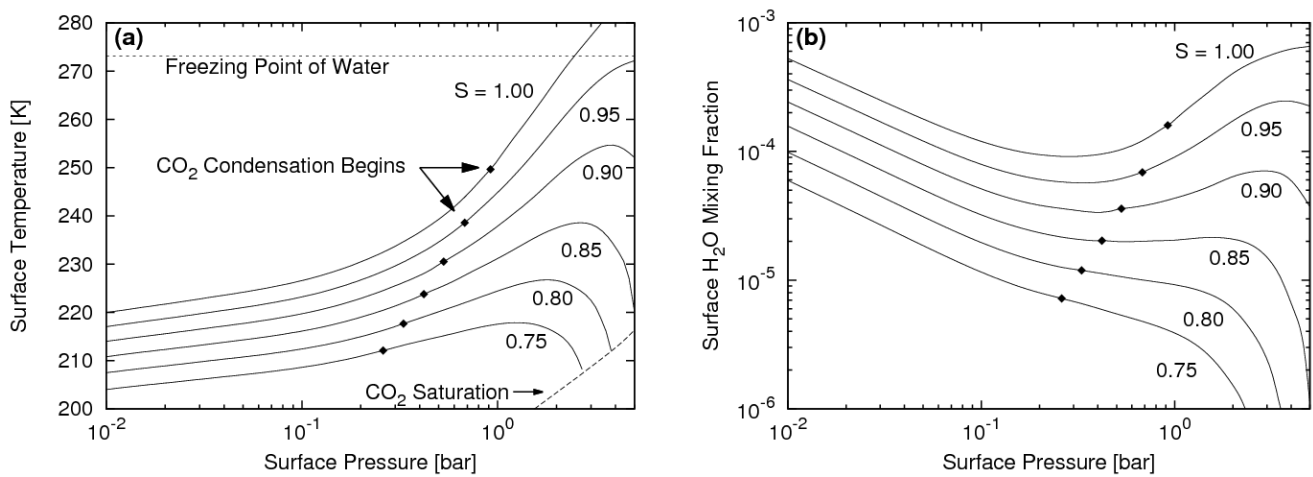

308  **Figure 2: Parameter space showing (a) surface temperature and (b) surface $H_2O$ mixing**

309  **fraction as a function of surface pressure *P* and insolation *S* (for modern Sun intensity *S* =**

310  **1.00). For each *S* track the pressure above which $CO_2$ condensation begins in the middle**

311  **atmosphere is indicated by ◆.**

312

313      To guide our search for atmospheres that would have been most conducive to the

314  formation of $O_3$, we first use the radiative-convective model with no photochemistry to explore

315  the parameter space of surface pressure *P* and insolation *S*, where *S* is the insolation as a fraction

316 of modern intensity. A plot of the resulting surface temperature is shown in Fig. 2a, with the
317 diamond symbols identifying for each value of $S$ at what surface pressure $CO_2$ condensation
318 begins in the middle atmosphere. Unlike the model of Kasting (1991), which produced $CO_2$
319 condensation at modern insolation levels for $P \geq 0.35$ bar, our model (without photochemistry)
320 does not condense until $P \geq 0.92$ bar. This discrepancy is likely due to the differing treatment of
321 NIR solar absorption between the models, with ours having greater stratospheric heating (J. F.
322 Kasting & R. Ramirez, pers. comm., 2011). It is worth noting that both models calculate a
323 globally averaged temperature profile and assume radiative-convective equilibrium. $CO_2$
324 condensation could still occur for $P < 0.92$ bar, as it does on Mars today, in cold regions at high
325 latitude (Kelly et al. 2006) and in the mesosphere due to temperature fluctuations driven by
326 atmospheric waves (Montmessin et al. 2006).

327     Surface $H_2O$ mixing fraction $f_{H2O}$ is shown in Fig. 2b as a function of $P$ and $S$. It may be
328 seen that for $S > 0.85$ there exists a local minimum for $f_{H2O}$. The reason for this minimum is as
329 follows: For $P < 0.3$ bar the surface temperature increases only weakly with pressure, causing the
330 vapor pressure of $H_2O$ to rise less quickly than the $CO_2$ pressure and resulting in the decline of
331 $f_{H2O}$. At $P > 0.3$ bar $H_2O$ becomes an important greenhouse gas for these atmospheres and its
332 positive feedback on temperature is strong enough to increase $f_{H2O}$ with increasing surface
333 pressure. For $S < 0.85$ the surface temperature never rises enough for $H_2O$ to become a
334 significant greenhouse gas, and so its mixing fraction decreases monotonically with $P$.

335     For the purpose of identifying conditions likely to support a thick $O_3$ layer, it is of interest
336 to summarize here some of the results of Rosenqvist & Chassefière (1995). Modeling a $CO_2$-$H_2O$
337 atmosphere with a fixed temperature profile and varying pressure, they identified three distinct

338  photochemical regimes for the $O_2$ mixing fraction $f_{O2}$, which may be compared with those for $f_{H2O}$

339  outlined above. For $P < 0.1$ bar $f_{O2}$ declined with $P$, while for $0.1 < P < 1$ bar it rose with $P$, and

340  finally declined again at $P > 1$ bar. The cause of rising $f_{O2}$ for 0.1 bar $< P < 1$ bar was determined

341  to be increased shielding of $H_2O$ against photolysis, resulting in decreased destruction of $O_2$ by

342  $HO_x$. By $P \sim 1$ bar this caused the atmospheric photochemistry to be insensitive to $H_2O$ content.

343  For P $< 1$ bar they found that artificially reducing $f_{H2O}$ (making the atmosphere drier) also

344  reduced $HO_x$ and produced a higher $f_{O2}$.

345      Taking the results presented above concerning pressure and temperature with no

346  photochemistry along with those of Rosenqvist & Chassefière (1995) concerning pressure at

347  fixed temperature, it would seem that conditions favorable to $O_3$ formation are most likely to be

348  found at moderate pressures of 0.3–1.0 bar. The lower end of this range corresponds with our

349  finding of a minimum mixing fraction of tropospheric $H_2O$ for pressures $\geq 0.3$ bar, while the

350  upper end is determined by the finding of Rosenqvist & Chassefière (1995) that tropospheric

351  $H_2O$ is increasingly shielded as pressures approach 1 bar. Thus, within this range we expect a

352  minimum of $H_2O$ photolysis, with a corresponding maximum in accumulation of oxidizing

353  species like $O_2$ and $O_3$.

354      In a broader context, focusing on atmospheric pressures of 0.3–1.0 bar is consistent with

355  recent work suggesting that Mars had an atmosphere $\geq 0.2$ bar at some time in its past (Manga et

356  al. 2012) but has volcanically outgassed $\leq 1$ bar $CO_2$ over its history (Grott et al. 2011). It is also

357  notable that this range falls within the 0.1–1 bar "dead zone" in Mars' history identified by

358  Richardson & Mischna (2005). That study pointed out that within this range of pressures diurnal

359  temperature highs would rarely reach the freezing point of water. Subsequently, loss of $CO_2$ by

360 surface weathering would be limited, reducing its utility for explaining the modern observed

361 pressure of 6 mbar. For climates in this "dead zone" we expect that feedback from $O_3$ would be

362 relatively strong. If a significant positive effect on surface temperature exists, warming by $O_3$

363 might have helped stabilize ephemeral liquid water, facilitating continued weathering out of $CO_2$.

364 We consider this possibility when investigating the effect of $O_3$ on the energy budget.

365

366 ### 3.3 Parameter Space Study With Photochemistry

367 Based on the results of the previous section, and in order to more directly compare with

368 previous work by Selsis et al. (2002), we select our nominal atmospheric model to have an $O_2$

369 mixing fraction $f_{O2} = 10^{-3}$ with surface pressure $P = 1$ bar and a modern insolation intensity $S =$

370 1.00. We then vary $f_{O2}$, $P$, and $S$ in turn to characterize their effect on the $O_3$ layer and its

371 feedbacks.

372

373 ### 3.3.1 $O_3$ Profile and Column

374 We begin our characterization of potential Martian $O_3$ layers by examining their

375 concentration profiles and integrated column densities. Varying $f_{O2}$, we find that with more $O_2$

376 the $O_3$ layer becomes thicker and its peak value larger and located at higher altitude (Fig. 3). This

377 same behavior has been found by models examining the photochemistry of early Earth (Kasting

378 & Donahue 1980), and is due to the increased availability of $O_2$ for $O_3$ production and decreased

379 penetration of UV. In contrast, we find that varying $P$ and $S$ have little effect on the profile when

380 altitude is plotted as pressure (not shown).

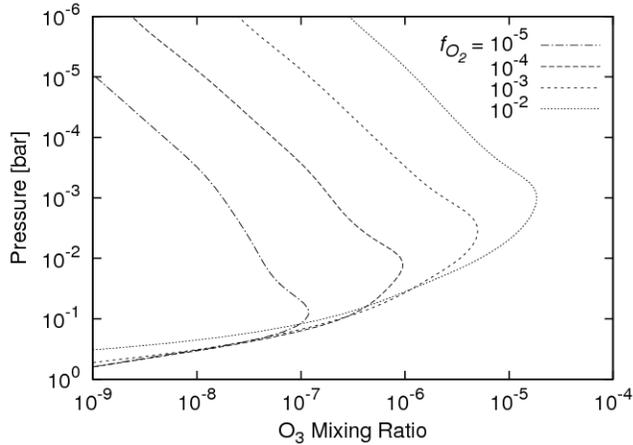

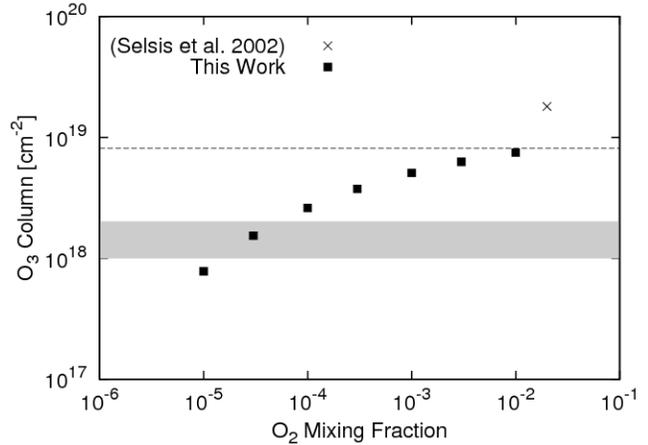

**Figure 3: O₃ profiles for various $f_{O2}$ in a 1 bar atmosphere under modern insolation.**

**Figure 4: O₃ column as a function of $f_{O2}$. The dashed line is a typical column on modern Earth ($8.1{\times}10^{18}$ cm⁻²), and the shaded region estimates the column necessary to protect bacteria from UV radiation (François & Gérard 1988).**

382    While increasing $f_{O2}$ generally increases the O₃ column (Fig. 4), it eventually levels off at

383    a value comparable to that of modern Earth (300 Dobson units $\approx 8.1{\times}10^{18}$ cm⁻²). Again, this is in

384    line with models of the early Earth (Kasting & Donahue 1980). We obtain an O₃ column of

385    $7.5{\times}10^{18}$ cm⁻² for our most oxidizing atmosphere with $f_{O2} = 1{\times}10^{-2}$. This is less than half of the

386    $1.8{\times}10^{19}$ cm⁻² column of the *"super ozone layer"* reported by Selsis et al. (2002) for a 1 bar

387    Martian atmosphere with $f_{O2} = 2{\times}10^{-2}$. While our results are not directly comparable with those

388    of Selsis et al. (2002) due to the different $f_{O2}$, as previously mentioned we expect our O₃ column

389    to be little different for higher $f_{O2}$. Curiously, since the model of Selsis et al. (2002) contains

390    catalytic HO$_x$ destruction of O₃ and ours does not, we would expect their model to produce a

391    column smaller than ours, not larger. Furthermore, both models contain the majority of their O₃

in the stratosphere, so the detailed tropospheric photochemistry implemented in Selsis et al. (2002) cannot be the cause of disagreement. We considered the possibility that the difference was due to our diurnal averaging of solar flux for photochemistry (Krasnopolsky 2006), but a test run using average daytime insolation increases our $O_3$ column by only 10%. The model's insensitivity to solar flux also rules out the disagreement being caused simply by a differing choice in solar zenith angle (Kasting & Donahue 1980). We conclude that the discrepancy most likely comes about because the result of Selsis et al. (2002) is not a converged solution, and should only be considered as an upper limit. Our simple converged model thus provides a tighter constraint on the maximum stratospheric $O_3$ column in a thick Martian atmosphere, limiting it to very nearly the same value as on Earth.

In the interest of assessing the potential habitability of these Martian atmospheres for primitive life, an estimate for the minimum $O_3$ column necessary to ensure bacterial survival from UV damage is included in Fig. 4. We follow Hiscox & Lindner (1997) in taking the minimum tolerable $O_3$ column range determined for Earth by François & Gérard (1988) and scaling it to account for the lower UV flux at Martian orbit. We find that, under this globally averaged, clear sky, 1 bar $CO_2$ atmosphere with modern insolation, surface UV conditions capable of allowing bacterial survival require $f_{O2} > 10^{-5}$. Including clouds and dust in the atmosphere would further reduce the UV flux at the surface. It should also be noted that in the early solar system, the UV flux from the young Sun was generally less than that of the modern Sun at $\lambda > 200$ nm and greater at $\lambda < 200$ nm (Ribas et al. 2010). Since $CO_2$ absorbs virtually all radiation at $\lambda < 200$ nm regardless of $O_2$ and $O_3$ column, the use of a modern solar spectrum by François & Gérard (1988) somewhat overestimates the minimum $O_3$ column required to support 414    primitive surface life on Mars under the young Sun.

415

416    <u>3.3.2 Energy Budget</u>

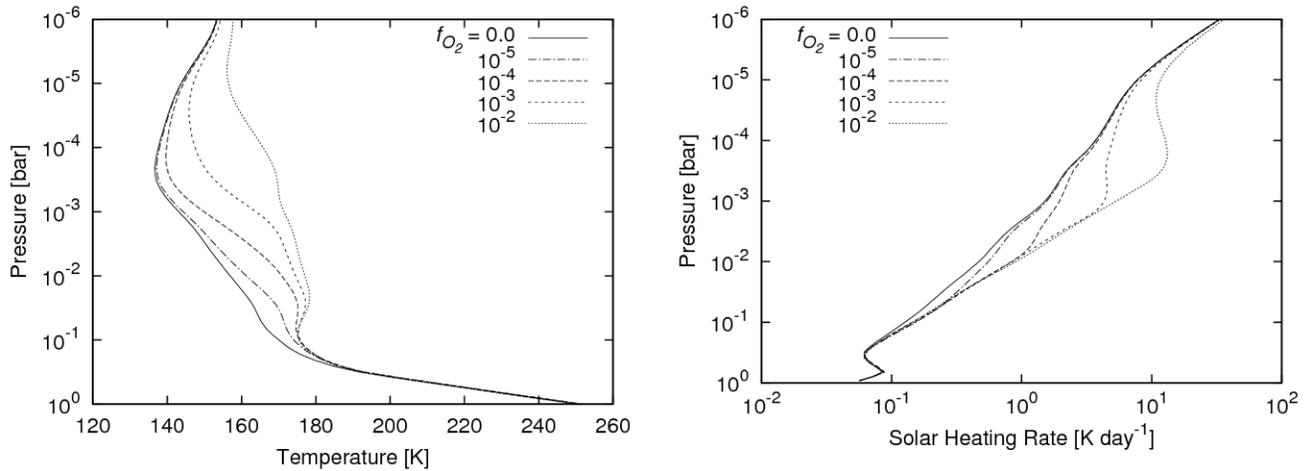

**Figure 5: Temperature profile for various $f_{O2}$ in a 1 bar atmosphere under modern insolation.**

**Figure 6: Solar heating profile for various $f_{O2}$ in a 1 bar atmosphere under modern insolation.**

417    While temperatures in the middle atmosphere strongly increase with more $O_2$ due to

418    increased solar absorption by the $O_3$ layer, the surface temperature is not greatly affected (Fig. 5).

419    Testing with the removal of the $O_3$ IR bands resulted in a surface temperature drop of at most ~ 1

420    K, indicating that the contribution of $O_3$ to the greenhouse effect in these models is weak. Solar

421    absorption by thin $O_3$ layers primarily cause heating just above the tropopause, and the altitude

422    and magnitude of peak solar heating by $O_3$ rises with increasing $O_2$ (Fig. 6). The majority of this

423    heating is due to absorption in the UV/VIS wavelengths, as on Earth, and occurs above the 0.1

424    bar pressure level in the atmosphere where Rayleigh scattering becomes important. As a result,

425    the planetary bond albedo is significantly reduced, with a decrease of 8.1% at $f_{O2} = 10^{-2}$ relative

426    to an atmosphere without an $O_3$ layer. This corresponds to a 3.8% increase in solar energy absorbed by the planet. The portion of this energy injected into the upper troposphere suppresses the thermal convective flux, resulting in a small warming effect at the surface. However, for thick $O_3$ layers this is counteracted by an overall reduction in the amount of solar radiation reaching the surface, as the majority of the $O_3$ absorption occurs high in the atmosphere and is simply re-radiated to space thermally. A listing of convective, longwave, and shortwave fluxes at the surface and top of the atmosphere for various values of $f_{O2}$ are given in Table 2.

| $f_{O_2}$ | $\Delta F^C_{\uparrow,surf}$ | $\Delta F^L_{\downarrow,surf}$ | $\Delta F^S_{\downarrow,surf}$ | $\Delta F^S_{\uparrow,top}$ |
|---|---|---|---|---|
| $10^{-5}$ | −0.3 (−0.6%) | +2.2 (+1.1%) | −0.6 (−0.6%) | −1.4 (−2.9%) |
| $10^{-4}$ | −0.6 (−1.2%) | +2.7 (+1.4%) | −1.3 (−1.3%) | −2.4 (−5.1%) |
| $10^{-3}$ | −1.6 (−3.2%) | −0.4 (−0.2%) | −1.9 (−2.0%) | −3.2 (−6.8%) |
| $10^{-2}$ | −2.3 (−4.8%) | −3.7 (−1.9%) | −2.4 (−2.5%) | −3.8 (−8.1%) |

**Table 2: Effect of varying $f_{O2}$ on $O_3$ energy budget forcing for $P$ = 1 bar, $S$ = 1.00. Each $\Delta F$ is taken with respect to the atmosphere with $f_{O2}$ = 0 and as has units of W/m². Superscripts C = convective, L = longwave, and S = shortwave.**

Holding $f_{O2} = 10^{-3}$ and lowering $S$ below 1.00 enhances the suppression of convection and alters the change in downward surface IR flux from a slight decrease to an increase (Table 3). In contrast, approximately constant relative reductions in the amount of solar radiation reaching the surface and reflecting to space are maintained, so the net effect of reducing $S$ is to enhance the relative surface warming effect of $O_3$. This suggests $O_3$ could act to buffer surface temperature against reductions in solar flux, although the effect is very weak for the model considered here.

| $S$ | $\Delta F^C_{\uparrow,surf}$ | $\Delta F^L_{\downarrow,surf}$ | $\Delta F^S_{\downarrow,surf}$ | $\Delta F^S_{\uparrow,top}$ |
|---|---|---|---|---|
| 1.00 | −1.6 (−3.2%) | −0.4 (−0.2%) | −1.9 (−2.0%) | −3.2 (−6.8%) |
| 0.95 | −1.4 (−3.3%) | +0.8 (+0.5%) | −1.9 (−2.0%) | −3.1 (−6.9%) |
| 0.90 | −1.3 (−3.5%) | +2.1 (+1.5%) | −1.9 (−2.1%) | −3.0 (−7.0%) |
| 0.85 | −1.2 (−3.8%) | +2.6 (+2.1%) | −1.9 (−2.1%) | −2.8 (−7.1%) |
| 0.80 | −1.2 (−4.8%) | +2.3 (+2.2%) | −1.7 (−2.1%) | −2.7 (−7.1%) |
| 0.75 | −1.3 (−6.2%) | +2.0 (+2.4%) | −1.6 (−2.1%) | −2.6 (−7.2%) |

**Table 3: Effect of varying $S$ on $O_3$ energy budget forcing for $P$ = 1 bar, $f_{O2} = 10^{-3}$. Each $\Delta F$ is taken with respect to the atmosphere with $f_{O2} = 0$ and as has units of W/m$^2$. Superscripts are the same as in Table 2.**

442

443  We also examine the effect of $O_3$ heating on $CO_2$ condensation in the middle atmosphere.
444 Fig. 7 shows that while a 1 bar atmosphere with $f_{O2} = 0$ at $S$ = 1.00 condenses near the 0.1 bar
445 level, an atmosphere with $f_{O2} = 10^{-3}$ does *not* due to heating by $O_3$. And though $CO_2$ condensation
446 is not completely suppressed at very low insolation ($S$ = 0.75), it does continue to be inhibited at
447 altitudes above the 0.1 bar level. This finding likely has ramifications for the ability of $CO_2$
448 clouds to warm the surface through scattering of IR (Forget & Pierrehumbert 1997; Mischna et
449 al. 2000), though accurately quantifying the effect would require detailed microphysical
450 modeling (Colaprete & Toon 2003).

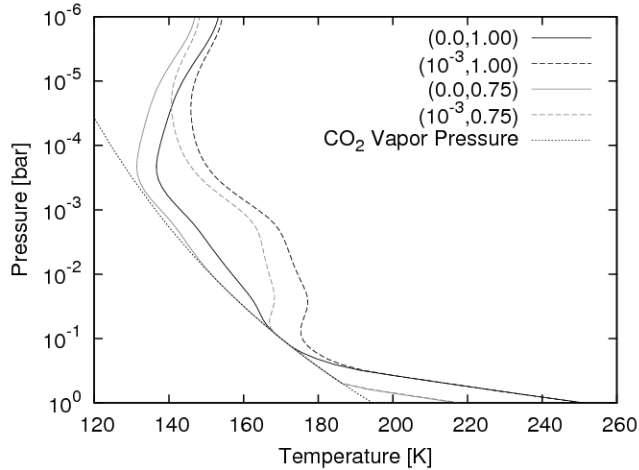

**Figure 7: Temperature profiles for $f_{O2} = 0$ and $1 \times 10^{-3}$ in a 1 bar atmosphere with $S = 1.00$ and 0.75. Line titles are in the format ($f_{O2}$, $S$).**

### 3.3.3 H₂O Dissociation

451

452     The effect of $O_3$ on the column integrated $H_2O$ dissociation rate is important because it

453     plays a role in both the atmosphere's photochemistry and its loss over time. In general, the

454     greater the amount of dissociation, the more the atmosphere will be dominated by the chemistry

455     of $HO_x$ products and the more $H_2$ will be produced. Increasing the concentration of $H_2$ in the

456     atmosphere directly increases its loss rate to space and thus the loss rate of $H_2O$ from Mars,

457     regardless of whether loss of $H_2$ is limited by the escape rate of H or diffusion of $H_2$ at the

458     homopause (R.E. Johnson et al. 2008; Zahnle et al. 2008). The concentration of $H_2$ in the upper

459     atmosphere can also indirectly affect the loss of other species by shielding them from the EUV

460     which powers their non-thermal escape mechanisms (Fox 2003).

461

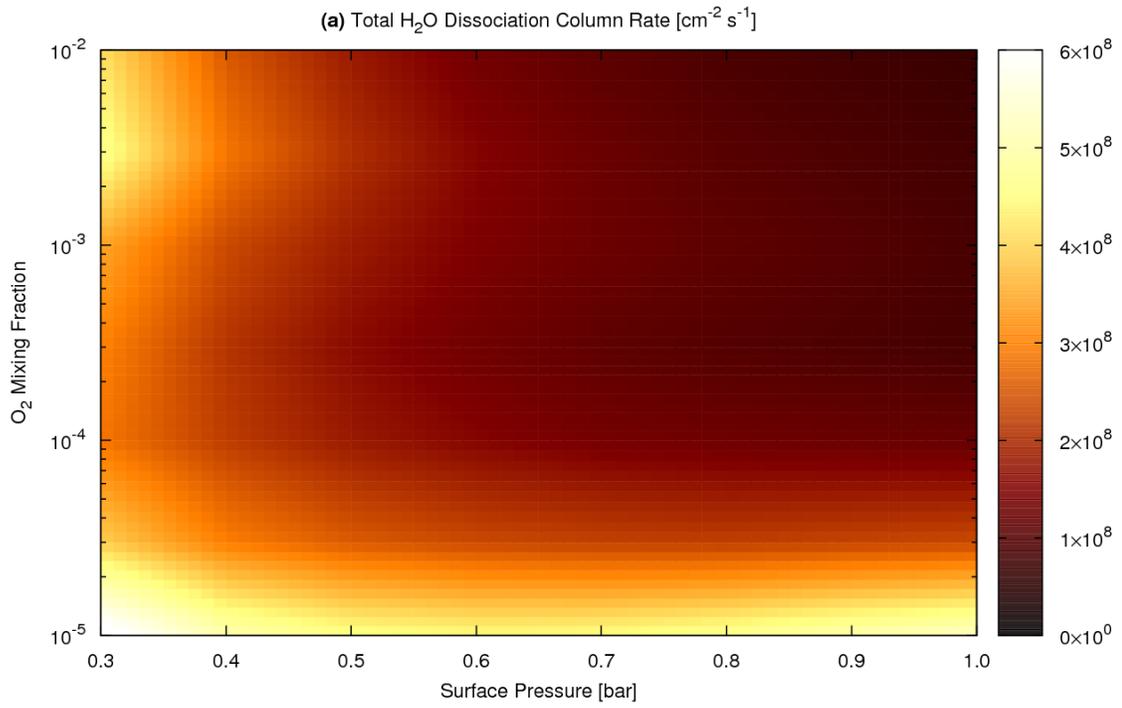

**(a)** Total H₂O Dissociation Column Rate [cm⁻² s⁻¹]

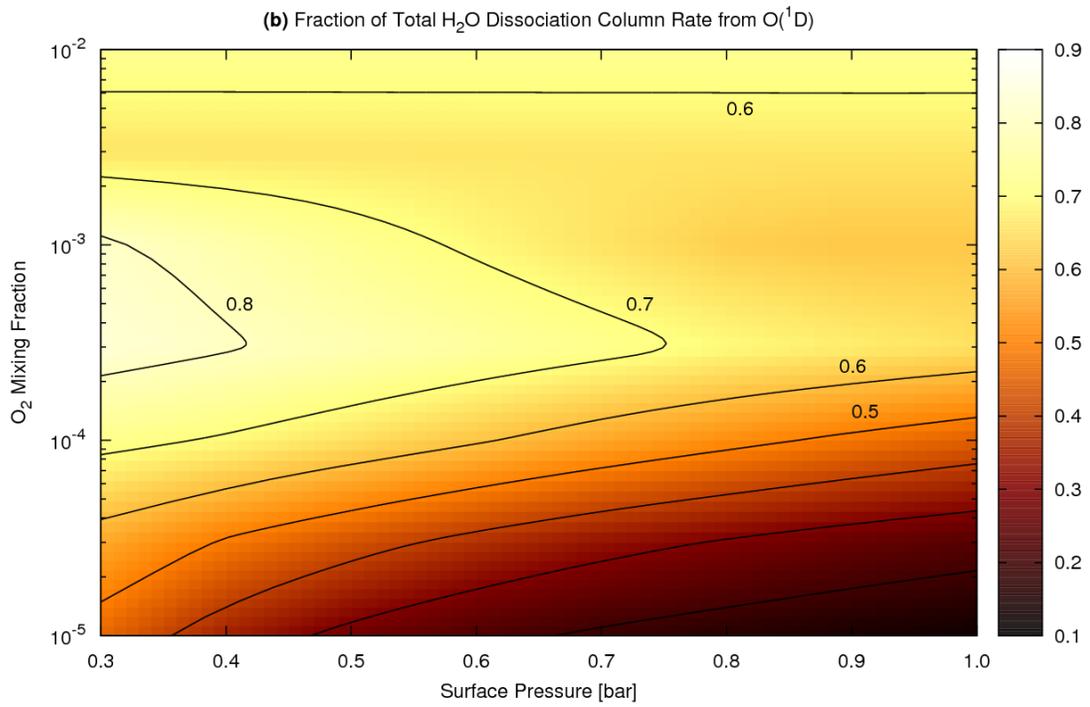

**(b)** Fraction of Total H₂O Dissociation Column Rate from O(¹D)

**Figure 8: Characterization of H₂O dissociation as a function of $f_{O2}$ and *P*: (a) total column integrated dissociation rate (photolysis + O($^1$D) attack) and (b) fraction of dissociation attributable to O($^1$D) attack.**

As seen in Table 1, the two pathways by which H₂O may be dissociated are direct photolysis by UV and attack by O($^1$D). The vast majority of O($^1$D) in our models was produced by photolysis of O₃ ($J_6$). In Fig. 8 we map the total column integrated H₂O dissociation rate and the fraction dissociated by O($^1$D) in the parameter space of *P* and $f_{O2}$. In low pressure atmospheres the column integrated O($^1$D) attack rate is large because the low altitude O₃ layer produces abundant O($^1$D) near the H₂O rich upper troposphere. However, for high pressure atmospheres the O₃ layer is raised up into the H₂O poor stratosphere, reducing the contribution of O($^1$D) attack to H₂O dissociation. For $f_{O2} \sim 10^{-5}$–$10^{-4}$ the O₃ layer provides little UV shielding, so the direct photolysis rate of H₂O is large at all surface pressures and tends to dominate dissociation. In the range $f_{O2} \sim 10^{-4}$–$10^{-3}$ a substantial O₃ layer accumulates and strongly suppresses H₂O photolysis. This lowers the total dissociation rate and causes O($^1$D) attack to dominate at all surface pressures. When $f_{O2} \sim 10^{-3}$–$10^{-2}$ a thick O₃ layer forms. At low surface pressure this strongly heats the cold trap, allowing more H₂O into the stratosphere where it can be directly photolyzed or attacked by O($^1$D). This enhancement in dissociation is not seen in high pressure atmospheres because the O₃ layer is high above the cold trap and warms it only weakly.

The details of these effects are illustrated for a 1 bar atmosphere in Fig. 9. We see that for low $f_{O2}$ most H₂O dissociation occurs by photolysis in the lower troposphere, similar to modern Mars (Nair et al. 1994). On the other hand, for high $f_{O2}$ dissociation by O($^1$D) attack becomes

485  dominant, as on modern Earth (Levy 1971). This is due to the shielding of direct $H_2O$ photolysis

486  by $O_3$ UV absorption and the enhanced production of $O(^1D)$ by photolysis of $O_3$. The shielding of

487  $H_2O$ is partially counteracted by humidification of the stratosphere, increasing the rate of $H_2O$

488  photolysis above the $O_3$ layer.

489

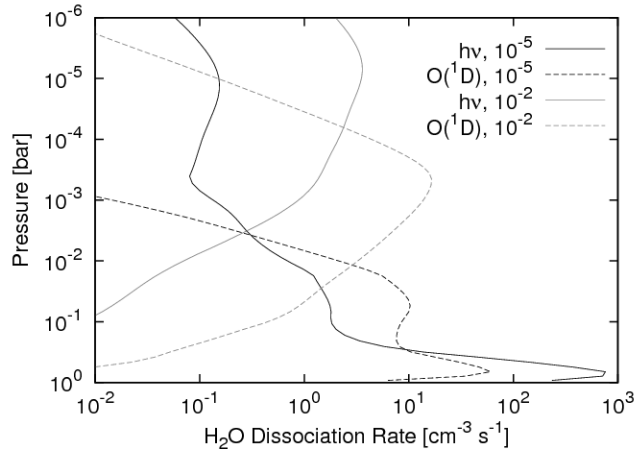

490  **Figure 9: Comparison of $H_2O$ dissociation profiles by photolysis and $O(^1D)$ attack for $f_{O2}$ =**

491  **$10^{-5}$ and $10^{-2}$. Line titles are in the format (dissociation type, $f_{O2}$).**

492

493  <u>3.3.4 Effects of $HO_x$ and Transport</u>

494       In the results presented here we have consciously omitted the effects of $HO_x$ chemistry

495  and atmospheric transport on the $O_3$ profile. These effects are especially important in the upper

496  troposphere, where the radiative forcing of $O_3$ is most sensitive to its concentration (Forster &

497  Shine 1997) and heating of the cold trap can influence stratospheric photochemistry. We expect

498  our model to overestimate the stratospheric $O_3$ concentration due to neglecting catalytic

499  destruction by $HO_x$ and downward transport to the troposphere. Conversely, we underestimate

500  the tropospheric $O_3$ concentration because of the neglect of downward transport from the
501  stratosphere, though this would be partially counteracted by $HO_x$ reactions. In terms of climate
502  forcing, this means that cooling by stratospheric $O_3$ is too large and warming by tropospheric $O_3$
503  too low. This suggests that the current results *underestimate* the climate warming effect of $O_3$
504  under the conditions examined.

505  In the results presented here, the peak $O_3$ concentration is always above the tropopause.
506  The location of this peak is determined by where the overhead column of $O_3$ absorbs most of the
507  UV capable of splitting $CO_2$ and $O_2$ to O, thus cutting off the means of $O_3$ production (Chapman
508  1930). Including catalytic $HO_x$ destruction would reduce the $O_3$ column at a given altitude in the
509  stratosphere, in turn lowering the altitude of peak $O_3$ concentration. This descent of the $O_3$ layer
510  into the upper troposphere would strengthen some of the feedback effects identified in this study,
511  such as inhibiting $CO_2$ condensation and enhancing $H_2O$ dissociation. Including convective
512  transport of $O_3$ from the stratosphere to the troposphere would have similar effects.

513  Taking into account the impact that $O_3$ can have on the $H_2O$ dissociation rate, and vice
514  versa, it is clear that a detailed model including $HO_x$ photochemistry and $O_3$ transport is needed
515  to more accurately characterize the feedback of an $O_3$ layer on Martian climate. Such
516  improvements will also allow us to calculate concentrations for $O_2$ and CO instead of using a
517  prescribed range of values, narrowing the parameter space to explore. This work is currently
518  underway.

519

520

521

## 4. Conclusions

We have explored here the possibility of a substantial $O_3$ layer accumulating in the Martian atmosphere for $O_2$ mixing fractions of $10^{-5}$ to $10^{-2}$, a range reasonable during the volcanically quiescent periods of Mars' history. Our study focused on atmospheres with surface pressures of 0.3–1.0 bar, as they are likely to have less $HO_x$ and thus more $O_2$ and $O_3$. We find that for modest $O_2$ mixing fractions of $f_{O2} > 10^{-5}$ an $O_3$ layer capable of effectively protecting primitive life from UV radiation would form. As $f_{O2}$ approaches $10^{-2}$ the $O_3$ column becomes comparable to that of modern Earth ($\sim 10^{19}$ cm$^{-2}$).

The presence of a thick $O_3$ layer decreases planetary albedo, up to 8.1% for a 1 bar atmosphere with $O_3$ column $\sim 10^{19}$ cm$^{-2}$. A portion of this extra heating goes toward inhibiting $CO_2$ condensation and weakening convection, both of which act to warm the surface for the clear sky conditions examined. However, for thick $O_3$ layers this surface warming is largely counteracted by stratospheric absorption of solar energy, which efficiently radiates back into space energy which otherwise would have reached the surface. Contribution of $O_3$ to the greenhouse effect is small under the conditions examined, with a surface temperature increase of only $\sim 1$ K. However, the net surface warming effect is likely underestimated due to our $O_3$ layers being biased toward higher altitudes. This results from the simplifications of neglecting $HO_x$ and eddy mixing in our photochemical model.

We also show that $O_3$ has a major influence on the column integrated dissociation rate of $H_2O$ by decreasing UV radiation in the lower atmosphere, increasing the production of $O(^1D)$, and increasing stratospheric humidity. Increasing the thickness of the $O_3$ layer causes $H_2O$ dissociation to shift from a "modern-Mars-like" state dominated by direct photolysis near the

544    surface to a "modern-Earth-like" state where photolysis is strongly suppressed and $O(^1D)$ attack

545    becomes dominant. For atmospheres > 0.5 bar, an increase in the thickness of the $O_3$ layer

546    generally lowered the $H_2O$ dissociation rate, with rates at $f_{O2} = 1 \times 10^{-2}$ being ~ $1/10^{th}$ their value

547    at $f_{O2} = 1 \times 10^{-5}$. This suggests the presence of a thick $O_3$ layer in Mars' past could have suppressed

548    the photochemical production of $H_2$, and, consequently, the loss of $H_2O$ to space. However, a

549    more complete photochemical model is required to make a definitive assessment of how a thick

550    $O_3$ layer would affect $H_2$ production on ancient Mars.

551

552    **Acknowledgments**

553    Thanks are given to D. Crisp, J. F. Kasting, and R. Ramirez for their helpful input on the

554    radiative-convective model. This work was supported by a grant from the NASA Planetary

555    Atmosphere program.

**References**

Alms, G.R., Burnham, A.K., Flygare, W.H., 1975. Measurement of the dispersion in polarizability anisotropies. J. Chem. Phys. 63, 3321–3326. doi:10.1063/1.431821

Baranov, Y.I., Lafferty, W.J., Fraser, G.T., 2004. Infrared spectrum of the continuum and dimer absorption in the vicinity of the $O_2$ vibrational fundamental in $O_2/CO_2$ mixtures. J. Mol. Spectr. 228, 432–440. doi:10.1016/j.jms.2004.04.010

Bideau-Mehu, A., Guern, Y., Abjean, R., Johannin-Gilles, A., 1973. Interferometric determination of the refractive index of carbon dioxide in the ultraviolet region. Opt. Commun. 9, 432–434. doi:10.1016/0030-4018(73)90289-7

Bougher, S.W., Blelly, P.L., Combi, M., Fox, J.L., Mueller-Wodarg, I., Ridley, A., Roble, R.G., 2008. Neutral Upper Atmosphere and Ionosphere Modeling. Space Science Reviews 139, 107–141. doi: 10.1007/s11214-008-9401-9.

Brown, L.R., Humphrey, C.M., Gamache, R.R., 2007. $CO_2$-broadened water in the pure rotation and $\nu_2$ fundamental regions. J. Mol. Spectr. 246, 1–21. doi:10.1016/j.jms.2007.07.010

Campbell, I.M., Gray, C.N., 1973. Rate constants for $O(^3P)$ recombination and association with $N(^4S)$. Chem. Phys. Let. 18, 607–609. doi:10.1016/0009-2614(73)80479-8

578

Carr, M., Head, J. W., 2010. Geologic history of Mars. Earth Planet. Sci. Lett. 294, 185-203. doi:10.1016/j.epsl.2009.06.042

581

Chapman, S., 1930. A Theory of Upper-Atmospheric Ozone. Memoirs of the Royal Meteorological Society 3, 103–125.

584

Clough, S.A., et al., 2005. Atmospheric radiative transfer modeling: a summary of the AER codes. JQSRT 91, 233–244. doi:10.1016/j.jqsrt.2004.05.058

587

Colaprete, A., Toon, O.B., 2003. Carbon dioxide clouds in an early dense Martian atmosphere. JGR 108, 5025. doi:10.1029/2002JE001967

590

Corrêa, M.P., Souza, R.A.F., Ceballos, J.C., Fomin, B., 2005. Preliminary results of simulations of a user-friendly fast line-by-line computer code for simulations of satellite signal. Anais XII Simpósio Brasileiro de Sensoriamento Remoto, Goiânia, Brasil, INPE, 363–370.

594

Daescu, D., Sandu, A., Carmichael, G.R., 2003. Direct and Adjoint Sensitivity Analysis of Chemical Kinetic Systems with KPP: II – Validation and Numerical Experiments, Atmospheric Environment 37, 5097–5114. doi:10.1016/j.atmosenv.2003.08.020

598

Damian, V., Sandu, A., Damian, M., Potra, F., Carmichael, G.R., 2002. The Kinetic PreProcessor

600   KPP – A Software Environment for Solving Chemical Kinetics. Computers and Chemical
601   Engineering 26, 1567–1579. doi:10.1016/S0098-1354(02)00128-X.

602

603   Eastman, E. D., 1929. Specific heats of gases at high temperatures. Tech. Paper 445, Bureau of
604   Mines.

605

606   Forget, F., Pierrehumbert, R.T., 1997. Warming early Mars with carbon dioxide clouds that
607   scatter infrared radiation. Science 278, 1273–1276. doi:10.1126/science.278.5341.1273

608

609   Forster, P. M. de F., K. P. Shine, 1997. Radiative forcing and temperature trends from
610   stratospheric ozone changes. JGR 102(D9), 10841–10855. doi:10.1029/96JD03510

611

612   Forster, P. M., et al., 2011. Evaluation of radiation scheme performance within chemistry climate
613   models. JGR 116, D10302. doi:10.1029/2010JD015361

614

615   Fox, J.L., 2003. Effect of $H_2$ on the Martian ionosphere: Implications for atmospheric evolution.
616   JGR 108(A6), 1223. doi:10.1029/2001JA000203

617

618   François, L.M., Gérard, J.C., 1988. Ozone, climate and biospheric environment in the ancient
619   oxygen-poor atmosphere. Planet. Space Sci. 36, 1391–1414. doi:0.1016/0032-
620   0633(88)90007-4

621

622     Grott, M., Morschhauser, A., Breuer, D., Hauber, E., 2011. Volcanic outgassing of $CO_2$ and $H_2O$
623         on Mars. Earth Planet. Sci. Lett. 308, 391–400. doi:10.1016/j.epsl.2011.06.014

624

625     Gruszka, M., Borysow, A., 1998. Computer simulation of the far infrared collision induced
626         absorption spectra of gaseous $CO_2$. Mol. Phys. 93, 1007–1016.
627         doi:10.1080/002689798168709

628

629     Halvey, I., Zuber, M.T., Schrag, D.P., 2007. A Sulfur Dioxide Climate Feedback on Early Mars.
630         Science 318, 1903–1907. doi:10.1126/science.1147039

631

632     Halevy, I., Pierrehumbert, R. T., and Schrag, D. P., 2009. Radiative transfer in $CO_2$-rich
633         paleoatmospheres. JGR 114, D18112. doi:10.1029/2009JD011915

634

635     Hauglustaine, D.A., Granier, C., Brasseur, G.P., Mégie, G., 1994. The importance of atmospheric
636         chemistry in the calculation of radiative forcing on the climate system. JGR 99(D1), 1173–
637         1186. doi:10.1029/93JD02987

638

639     Hiscox, J.A., B.L. Lindner, 1997. Ozone and planetary habitability. Journal of the British
640         Interplanetary Society 50, 109–114.

641

642     Hunten, D., 1993. Atmospheric Evolution of the Terrestrial Planets. Science 259, 915–920.
643         doi:10.1126/science.259.5097.915

644

Inn, E.C.Y., 1974. Rate of recombination of oxygen atoms and CO at temperatures below ambient. J. Chem. Phys. 61, 1589–1590. doi:10.1063/1.1682139

647

Johnson, R.E., Combi, M.R., Fox, J.L., Ip, W.H., Leblanc, F., McGrath, M.A., Shematovich, V.I., Strobel, D.F., Waite, J.H., 2008. Exospheres and Atmospheric Escape. Space Science Reviews 139, 355–397. doi:10.1007/s11214-008-9415-3

651

Johnson, S.S., Mischna, M.A., Grove, T.L., Zuber, M.T., 2008. Sulfur-induced greenhouse warming on Early Mars. JGR 113, E08005. doi:10.1029/2007JE002962

654

Johnson, S.S., Pavlov, A.A., Mischna, M.A., 2009. Fate of $SO_2$ in the ancient Martian atmosphere: Implications for transient greenhouse warming. JGR 114, E11011. doi:10.1029/2008JE003313

658

Kasting, J.F., Liu, S.C., Donahue, T.M., 1979. Oxygen Levels in the Prebiological Atmosphere. JGR 84(C6), 3097–3107. doi:10.1029/JC084iC06p03097

661

Kasting, J.F., Donahue, T.M., 1980. The Evolution of Atmospheric Ozone. JGR 85(C6), 3255–3263. doi:10.1029/JC085iC06p03255

664

Kasting, J. F., Walker, J.C.G., 1981. Limits on Oxygen Concentration in the Prebiological

Atmosphere and the Rate of Abiotic Fixation of Nitrogen. JGR 86(C2), 1147–1158. doi:10.1029/JC086iC02p01147

Kasting, J. F., Holland, H.D., Pinto, J.P., 1985. Oxidant Abundances in Rainwater and the Evolution of Atmospheric Oxygen. JGR 90(D6), 10497–10510. doi:10.1029/JD090iD06p10497

Kasting, J. F., 1991. $CO_2$ Condensation and the Climate of Early Mars. Icarus 94, 1–13. doi:10.1016/0019-1035(91)90137-I

Kasting, J.F., 1995. $O_2$ concentrations in dense primitive atmospheres: commentary. Planet. Space Sci. 43, 11–13. doi:10.1016/0032-0633(94)00203-4

Kasting, J.F., 1997. Warming Early Earth and Mars. Science 276, 1213–1215. doi:10.1126/science.276.5316.1213

Kelly, N. J., et al., 2006. Seasonal polar carbon dioxide frost on Mars: $CO_2$ mass and columnar thickness distribution. JGR 111, E03S07. doi:10.1029/2006JE002678

Kiehl, J.T., Trenberth, K.E., 1997. Earth's Annual Global Mean Energy Budget. Bull. Amer. Meteor. Soc., 78, 197-208. doi:10.1175/1520-0477(1997)078%3C0197:EAGMEB%3E2.0.CO;2

688

Krasnopolsky, V.A., 2006. Photochemistry of the martian atmosphere: Seasonal, latitudinal, and diurnal variations. Icarus 185, 153–170. doi:10.1016/j.icarus.2006.06.003

Kuntz, M., Höpfner, M., 1999. Efficient line-by-line calculation of absorption coefficients. JQSRT 63, 97–114. doi:10.1016/S0022-4073(98)00140-X

Kurucz R.L., 1992. Synthetic infrared spectra in infrared solar physics. In: Rabin D.M., Jefferies J.T., (Eds.), IAU Symposium, 154. Kluwer Academic Press, Norwell, MA.

Lebonnois, S., Hourdin, F., Eymet, V., Crespin, A., Fournier, R., Forget, F., 2010. Superrotation of Venus' atmosphere analyzed with a full general circulation model. JGR 115, E06006. doi:10.1029/2009JE003458.

Lemmon, E.W., Huber, M.L., McLinden, M.O., 2010. NIST Standard Reference Database 23: Reference Fluid Thermodynamic and Transport Properties-REFPROP, Version 9.0, National Institute of Standards and Technology, Standard Reference Data Program, Gaithersburg.

Letchworth, K., Benner, D., 2007. Rapid and accurate calculation of the Voigt function. JQSRT 107, 173–192. doi:10.1016/j.jqsrt.2007.01.052

709 Levy, H., 1971. Normal Atmosphere: Large Radical and Formaldehyde Concentrations

710      Predicted. Science 173, 141–143. doi:10.1126/science.173.3992.141

711

712 Li, H., Ji, X., Yan, J., 2006. A new modification on RK EOS for gaseous CO2and gaseous

713      mixtures of $CO_2$ and $H_2O$. Int. J. Energy Res. 30, 135–148. doi:10.1002/er.1129

714

715 Liou, KN., 2002. An introduction to atmospheric radiation. 2nd ed., Academic Press. p. 126

716

717 López-Valverde, M., Edwards, D., López-Puertas, M., Roldán, C., 1998. Non-local

718      thermodynamic equilibrium in general circulation models of the Martian atmosphere 1.

719      Effects of the local thermodynamic equilibrium approximation on thermal cooling and solar

720      heating. JGR 103, 16799–16811. doi:10.1029/98JE01601

721

722 Manga, M., Patel, A., Dufek, J., Kite, E.S., 2012. Wet surface and dense atmosphere on early

723      Mars suggested by the bomb sag at Home Plate Mars. Geophys. Res. Lett. 39, L01202.

724      doi:10.1029/2011GL050192

725

726 McElroy, M.B., Donahue, T.M., 1972. Stability of the Martian atmosphere. Science 177, 986–

727      988. doi:10.1126/science.177.4053.986

728

729    Meadows, V. S., Crisp, D., 1996. Ground-based near-infrared observations of the Venus

730        nightside: The thermal structure and water abundance near the surface. JGR 101(E2), 4595–

731        4622. doi:10.1029/95JE03567

732

733    Minschwaner, K., Anderson, G.P., Hall, L.A., Yoshino, K. 1992. Polynomial coefficients for

734        calculating $O_2$ Schumann-Runge cross sections at 0.5 cm$^{-1}$ resolution. JGR 97(D9), 10103–

735        10108. doi: 10.1029/92JD00661

736

737    Mischna, M.A., Kasting, J.F., Pavlov, A., Freedman, R., 2000. Influence of carbon dioxide

738        clouds on early martian climate. Icarus 145, 546–554. doi:10.1006/icar.2000.6380

739

740    Montmessin, F., et al., 2006. Subvisible $CO_2$ ice clouds detected in the mesosphere of Mars.

741        Icarus 183, 403–410. doi:10.1016/j.icarus.2006.03.015

742

743    Montmessin, F., et al., 2011. A layer of ozone detected in the nightside upper atmosphere of

744        Venus. Icarus 216, 82–85. doi:10.1016/j.icarus.2011.08.010

745

746    Nair, H., Allen, M., Anbar, A.D., Yung, Y.L., Clancy, R.T., 1994. A Photochemical Model of the

747        Martian Atmosphere. Icarus 111, 124–150. doi:10.1006/icar.1994.1137

748

749    Neukum , G., et al., 2010. The geologic evolution of Mars: Episodicity of resurfacing events and

750        ages from cratering analysis of image data and correlation with radiometric ages of Martian

751       meteorites. Earth Planet. Sci. Lett. 294, 204–222. doi:10.1016/j.epsl.2009.09.006

752

753     Old, J.G., Gentili, K.L., Peck, E.R., 1971. Dispersion of Carbon Dioxide. J. Opt. Soc. Am. 61,

754       89–90. doi:10.1364/JOSA.61.000089

755

756     Parkinson, T.D., Hunten, D.M., 1972. Spectroscopy and Aeronomy of $O_2$ on Mars. J. Atm. Sci.

757       29, 1380–1390. doi:10.1175/1520-0469(1972)029%3C1380:SAAOOO%3E2.0.CO;2

758

759     Perrin, M.Y., Hartmann, J.M., 1989. Temperature-dependent measurements and modeling of

760       absorption by $CO_2$–$N_2$ mixtures in the far line-wings of the 4.3-μm $CO_2$ band. JQSRT 42,

761       311–317. doi:10.1016/0022-4073(89)90077-0

762

763     Pollack, J.B., Kasting, J.F., Rochardson, S.M., Poliakoff, K., 1987. The case for a wet, warm

764       climate on early Mars. Icarus 71, 203–224. doi:10.1016/0019-1035(87)90147-3

765

766     Quine, B.M., and Drummond, J.R., 2002. GENSPECT: a line-by-line code with selectable

767       interpolation error tolerance. JQSRT 74, 147–165. doi:10.1016/S0022-4073(01)00193-5

768

769     Ribas, I., et al., 2010. Evolution of the Solar Activity Over Time and Effects on Planetary

770       Atmospheres. II. $\kappa^1$ Ceti, an Analog of the Sun when Life Arose on Earth. Astrophys. J. 714,

771       384–395. doi:10.1088/0004-637X/714/1/384

772

773    Richardson, M.I., Mischna, M.A., 2005. Long-term evolution of transient liquid water on Mars.
774        JGR 110, E03003. doi:10.1029/2004JE002367

775

776    Robbins, S.J., Achille, G.D., Hynek, B.M., 2011. The volcanic history of Mars: High-resolution
777        crater-based studies of the calderas of 20 volcanoes. Icarus 211, 1179–1203.
778        doi:10.1016/j.icarus.2010.11.012

779

780    Rothman, L.S., et al., 2009. The HITRAN 2008 molecular spectroscopic database. JQSRT 110,
781        533–572. doi:10.1016/j.jqsrt.2009.02.013

782

783    Rosenqvist, J., Chassefière, E., 1995. Inorganic chemistry of $O_2$ in a dense primitive atmosphere.
784        Planet. Space Sci. 43, 3–10. doi:10.1016/0032-0633(94)00202-3

785

786    Sander S.P., et al., 2011. Chemical Kinetics and Photochemical Data for Use in Atmospheric
787        Studies, Evaluation Number 17, JPL Publication 10-6, Jet Propulsion Laboratory, Pasadena,
788        CA.

789

790    Sandu, A., Daescu, D., Carmichael,G.R., 2003. Direct and Adjoint Sensitivity Analysis of
791        Chemical Kinetic Systems with KPP: I – Theory and Software Tools, Atmospheric
792        Environment 37, 5083–5096. doi:10.1016/j.atmosenv.2003.08.019

793

794    Segura, A., Meadows, V. S., Kasting, J. F., Crisp, D., Cohen, M., 2007. Abiotic formation of $O_2$

795      and $O_3$ in high-$CO_2$ terrestrial atmospheres. A&A 472, 665–679. doi:[10.1051/0004-](#)

796      [6361:20066663](#)

797

798    Selsis, F., Despois, D., Parisot, J.-P., 2002. Signature of life on exoplanets: Can Darwin produce

799      false positive detections? A&A 388, 985–1003. doi:[10.1051/0004-6361:20020527](#)

800

801    Span, R., Wagner, W., J., 1996. A New Equation of State for Carbon Dioxide Covering the Fluid

802      Region from the Triple-Point Temperature to 1100 K at Pressures up to 800 MPa. Phys.

803      Chem. Ref. Data 25, 1509–11596. doi:[10.1063/1.555991](#)

804

805    Squyres, S.W., Kasting, J.F., 1994. Early Mars: How Warm and How Wet? Science 265, 744–

806      749. doi:[10.1126/science.265.5173.744](#)

807

808    Tian, F., Claire, M.W., Haqq-Misra, J.D., Smith, M., Crisp, D.C., Catling, D., Zahnle, K.,

809      Kasting, J.F., 2010. Photochemical and climate consequences of sulfur outgassing on early

810      Mars. Earth Planet. Sci. Lett. 295, 412–418. doi:[10.1016/j.epsl.2010.04.016](#)

811

812    Tonkov, M.V., et al., 1996. Measurements and empirical modeling of pure $CO_2$ absorption in the

813      2.3-µm region at room temperature: Far wings, allowed and collision-induced bands. Appl.

814      Opt. 35, 4863–4870. doi:[10.1364/AO.35.004863](#)

815

816    Toon, O.B., McKay, C.P., Ackerman, T.P., Santhanam, K., 1989. Rapid calculation of radiative

817    heating rates and photodissociation rates in inhomogeneous multiple scattering atmospheres.

818    JGR 94(D13), 16287–16301. doi:10.1029/JD094iD13p16287

819

820    Tvorogov, S.D, Rodimova, O.B., Nesmelova, L.I., 2005. On the correlated *k*-distribution

821    approximation in atmospheric calculations. Opt. Eng. 44, 071202. doi:10.1117/1.1955318

822

823    Wells, R.J., 1999. Rapid approximation to the Voigt/Faddeeva function and its derivatives.

824    JQSRT 64, 29–48. doi:10.1016/S0022-4073(97)00231-8

825

826    Wordsworth, R., Forget, F., Eymet, V., 2010a. Infra-red collision-induced and far-line absorption

827    in dense CO2 atmospheres. Icarus 210, 992–997. doi:10.1016/j.icarus.2010.06.010

828

829    Wordsworth, R., Forget, F., Selsis, F., Madeleine, J.-B., Millour, E., Eymet, V., 2010b. Is Gliese

830    581d habitable? Some constraints from radiative-convective climate modeling. A&A 522,

831    A22. doi:10.1051/0004-6361/201015053

832

833    Yung Y. L., Nair H., Gerstell M. F., 1997. $CO_2$ Greenhouse in the Early Martian Atmosphere:

834    $SO_2$ Inhibits Condensation. Icarus 130, 222–224. doi:10.1006/icar.1997.5808

835

836    Zahnle, K. J., Haberle, R. M., Catling, D. C., Kasting, J. F., 2008. Photochemical instability of

837    the ancient Martian atmosphere. JGR 113, E11004. doi:10.1029/2008JE003160

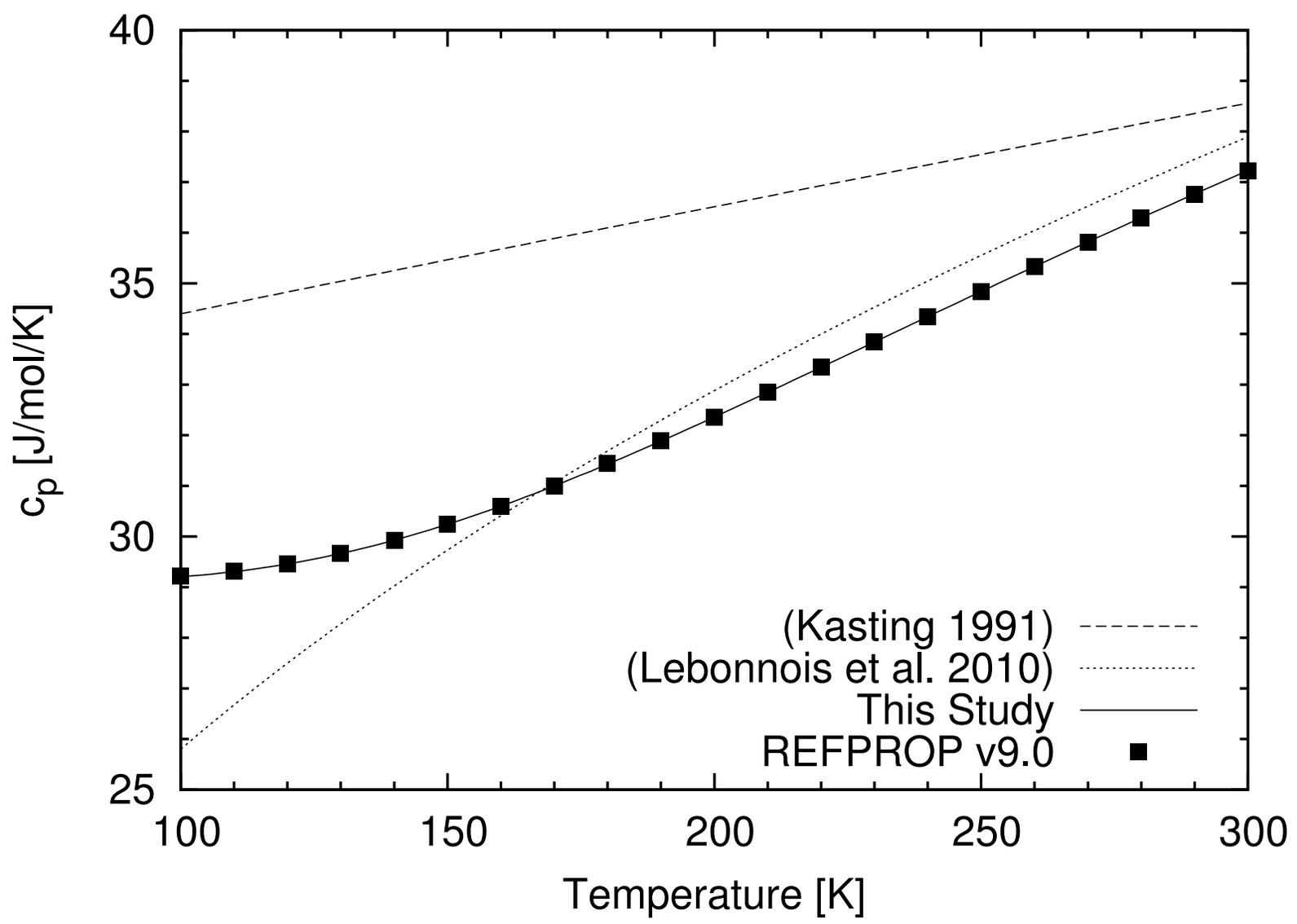

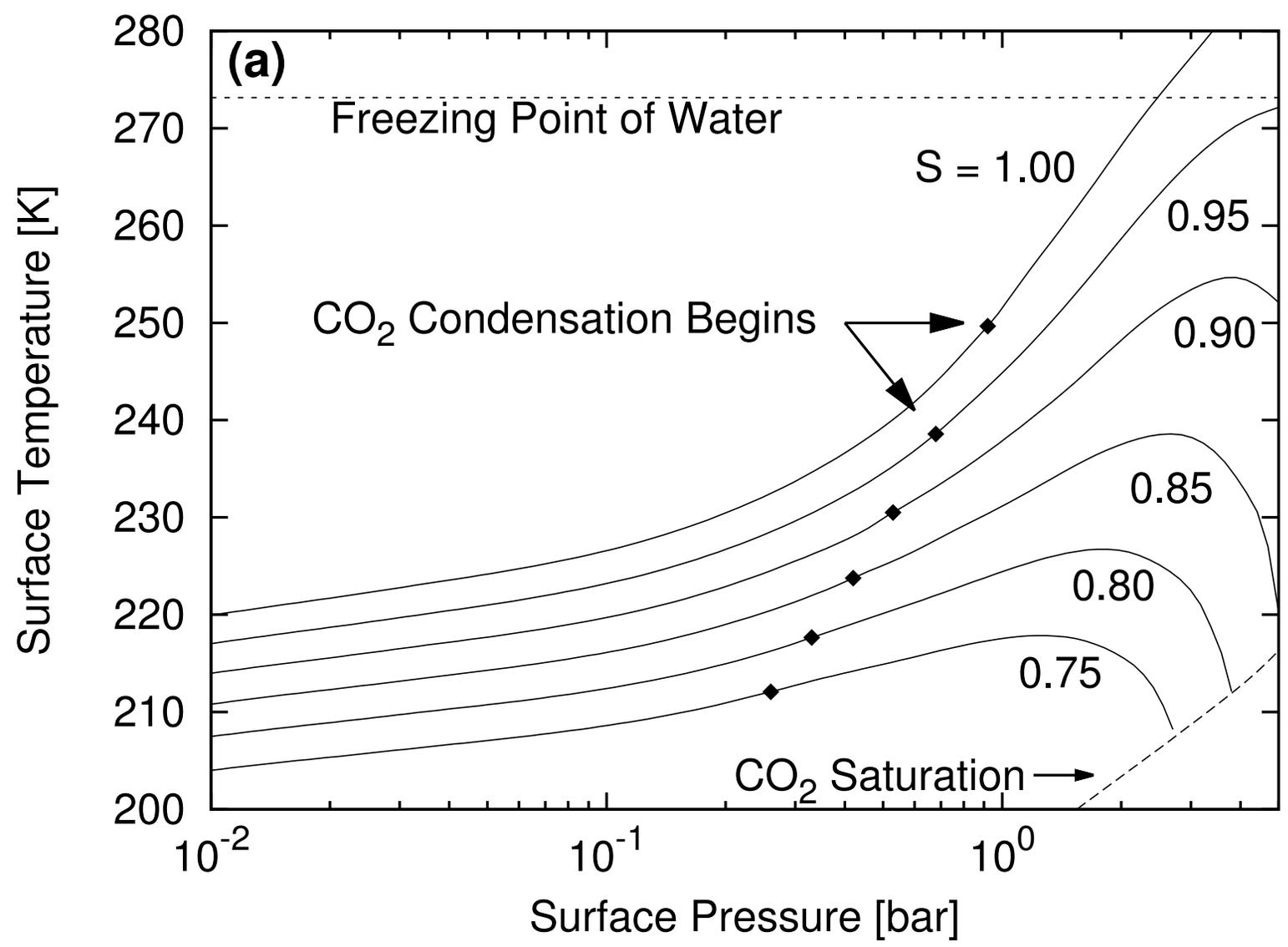

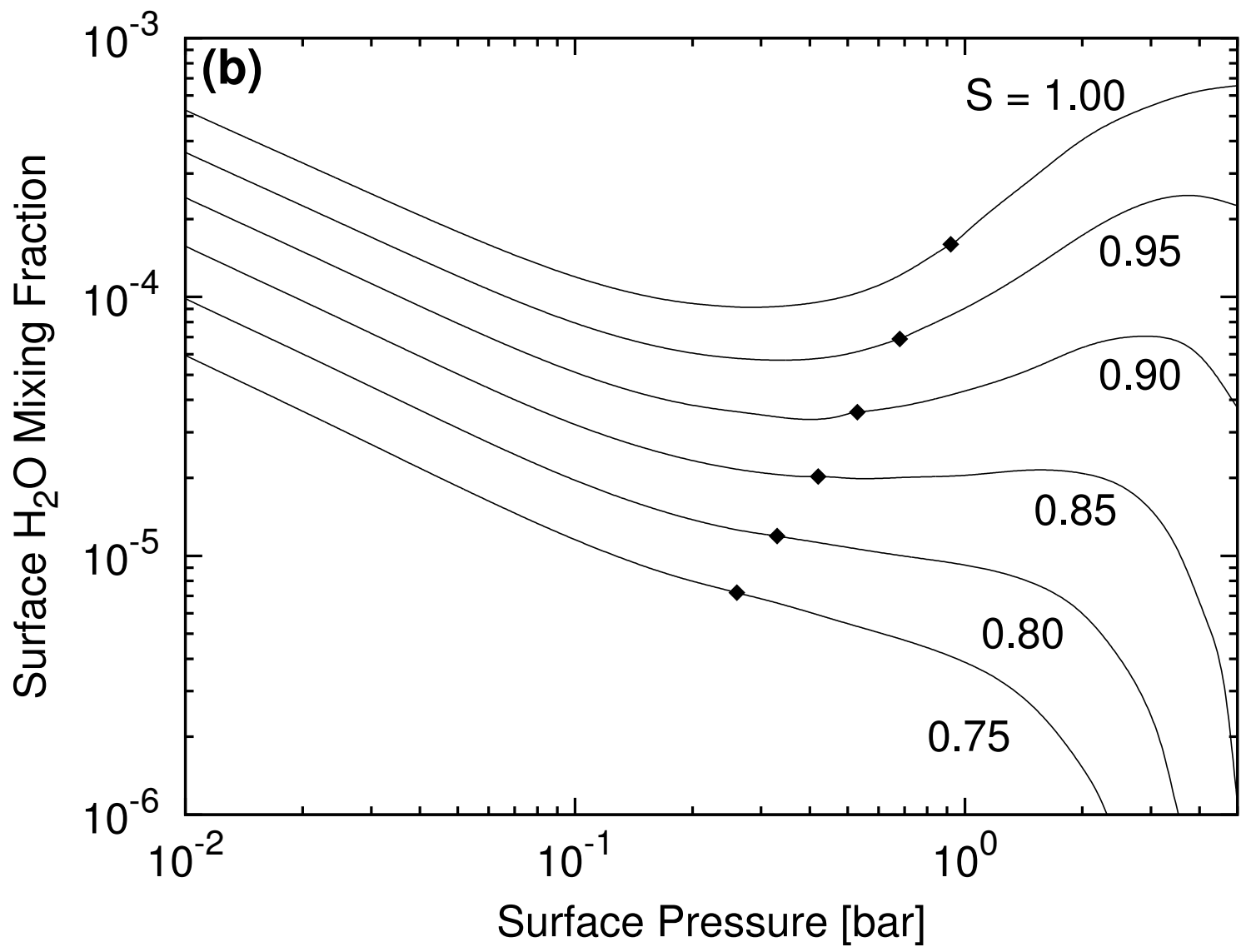

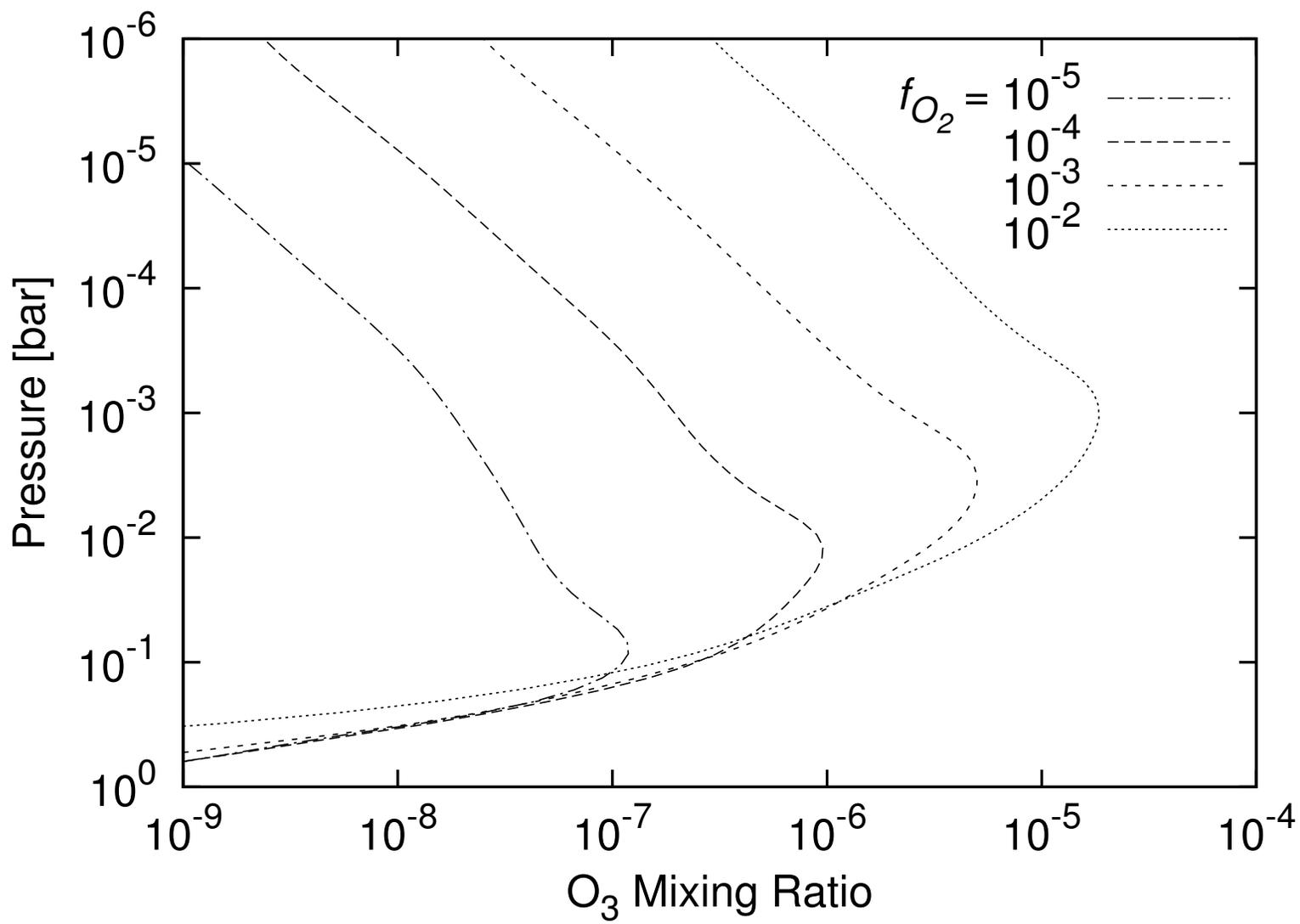

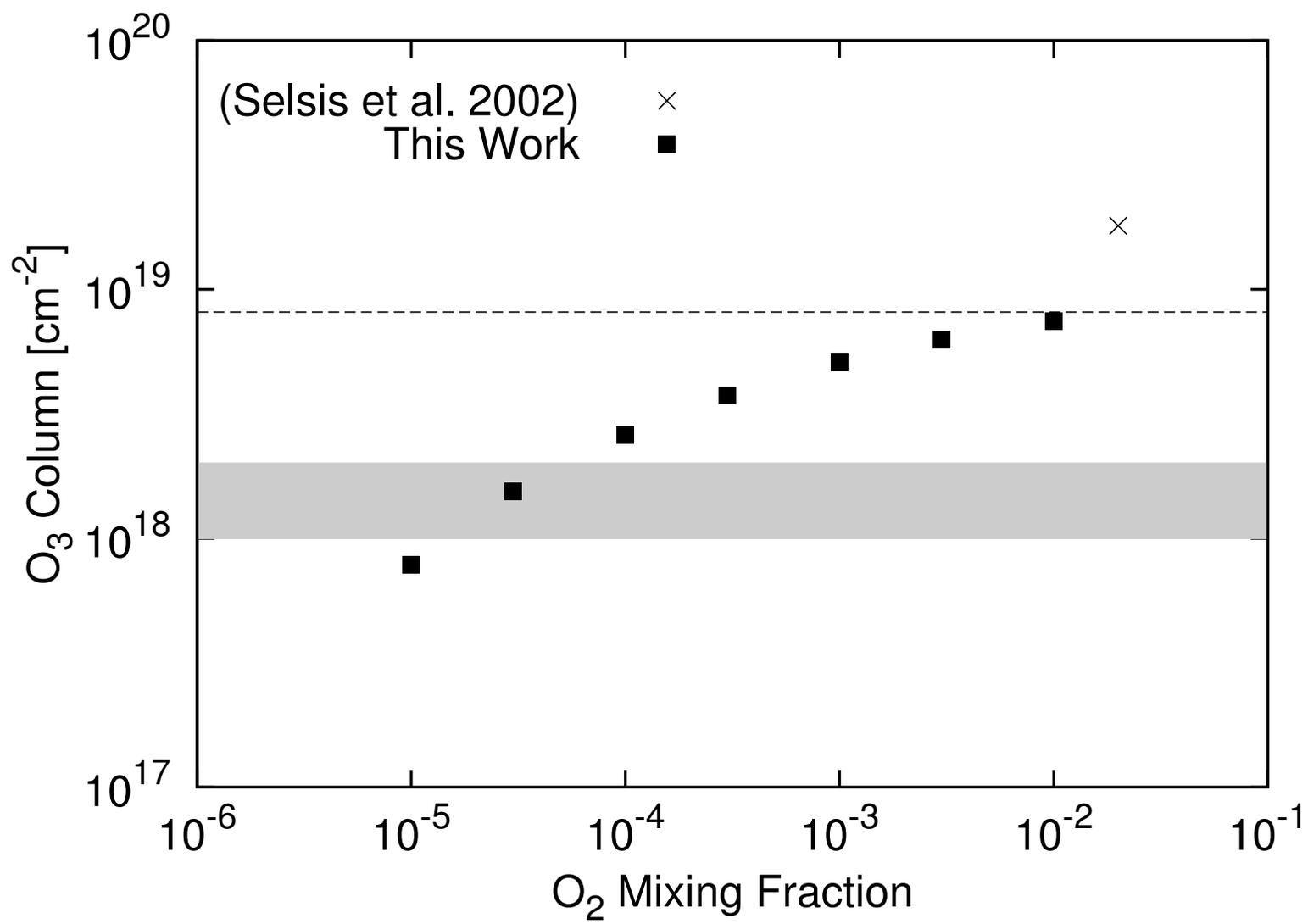

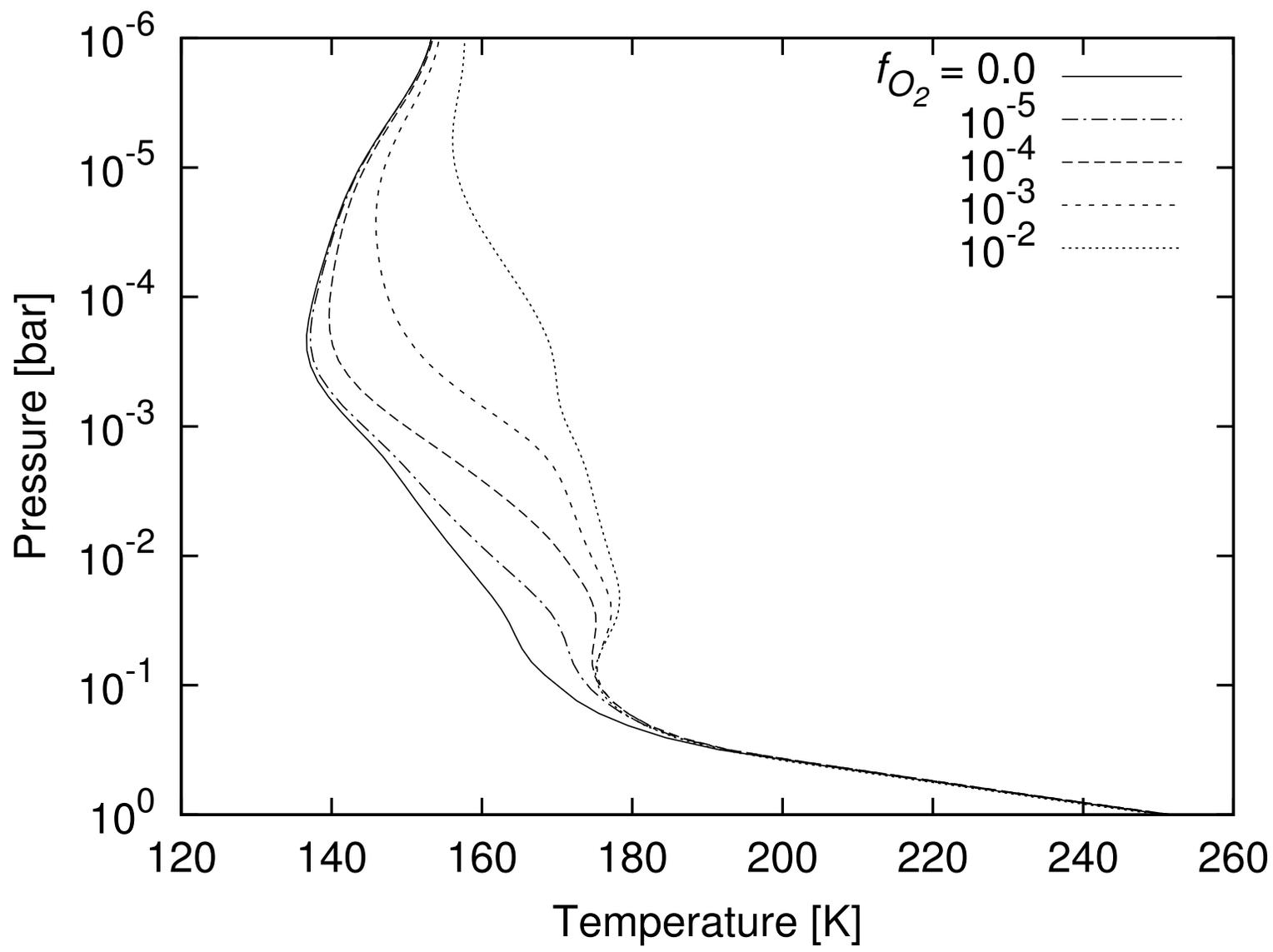

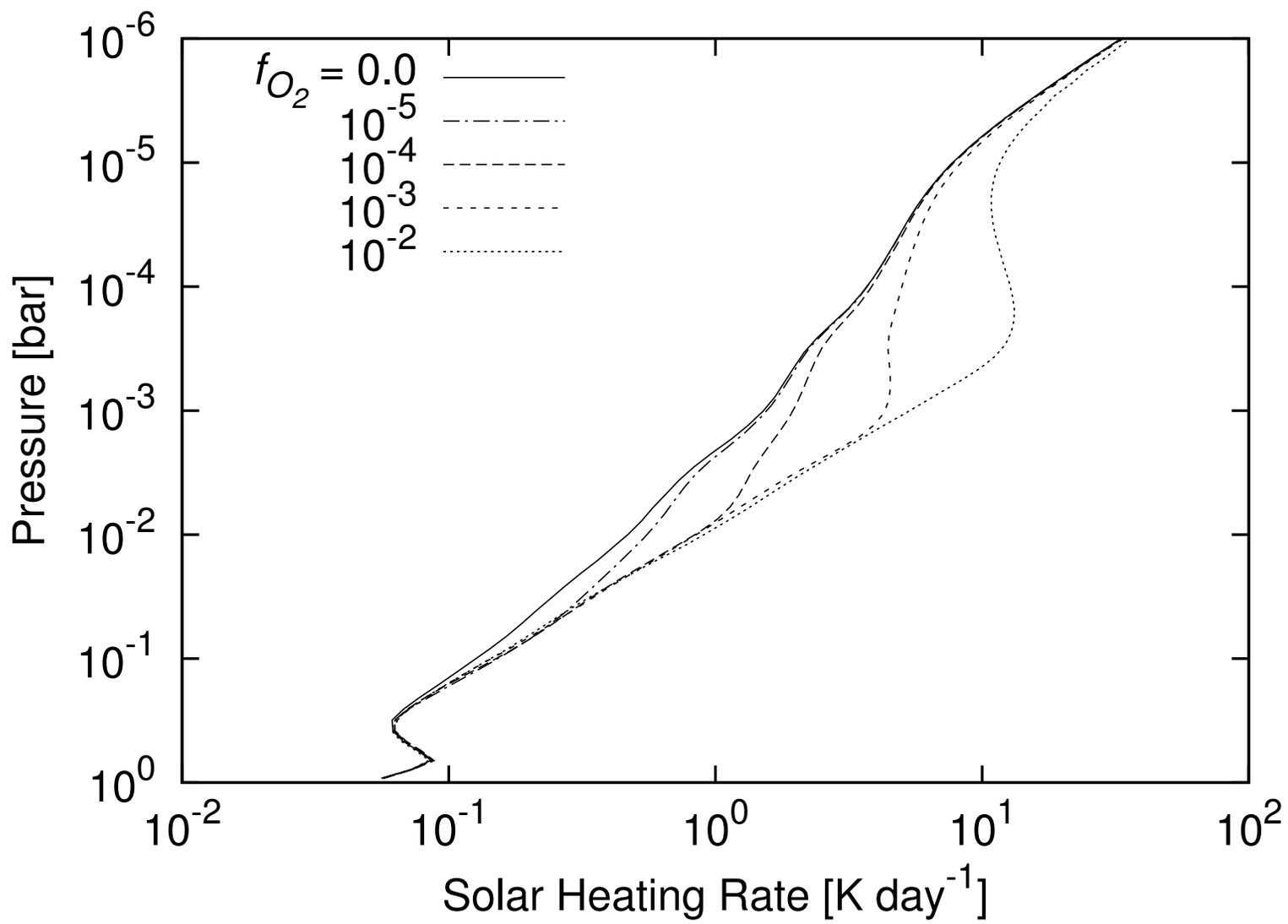

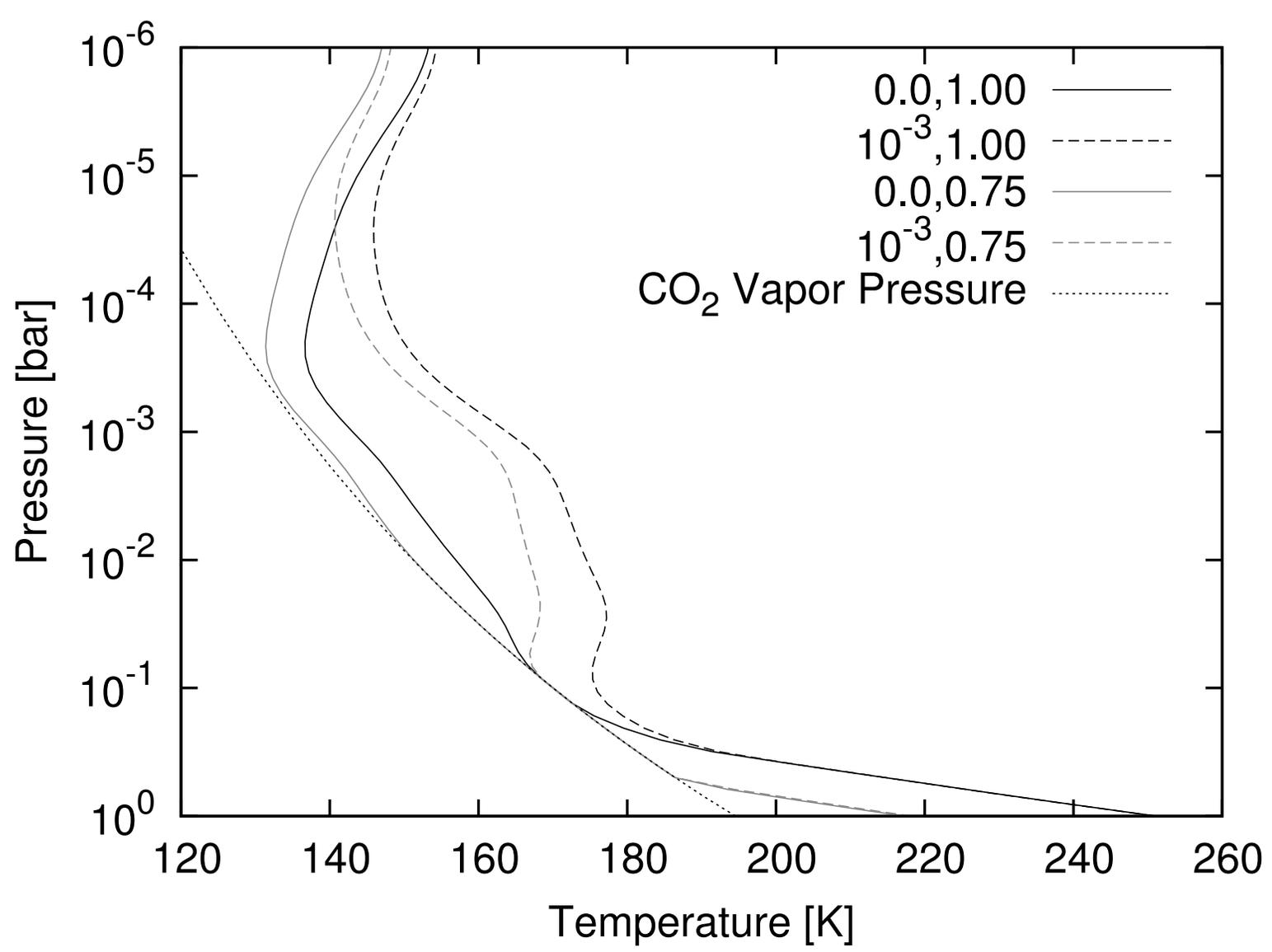

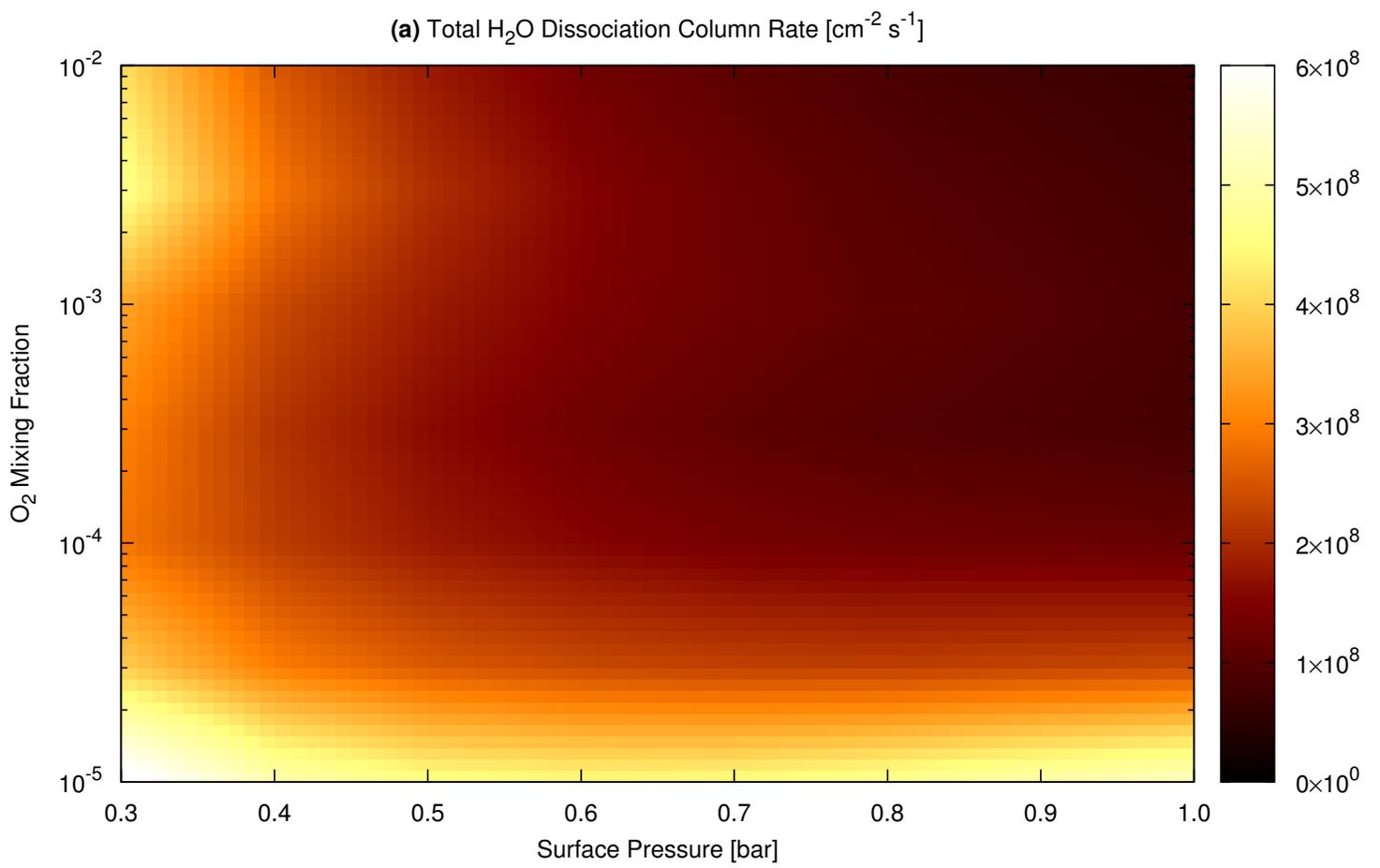

**Figure8a Grey**

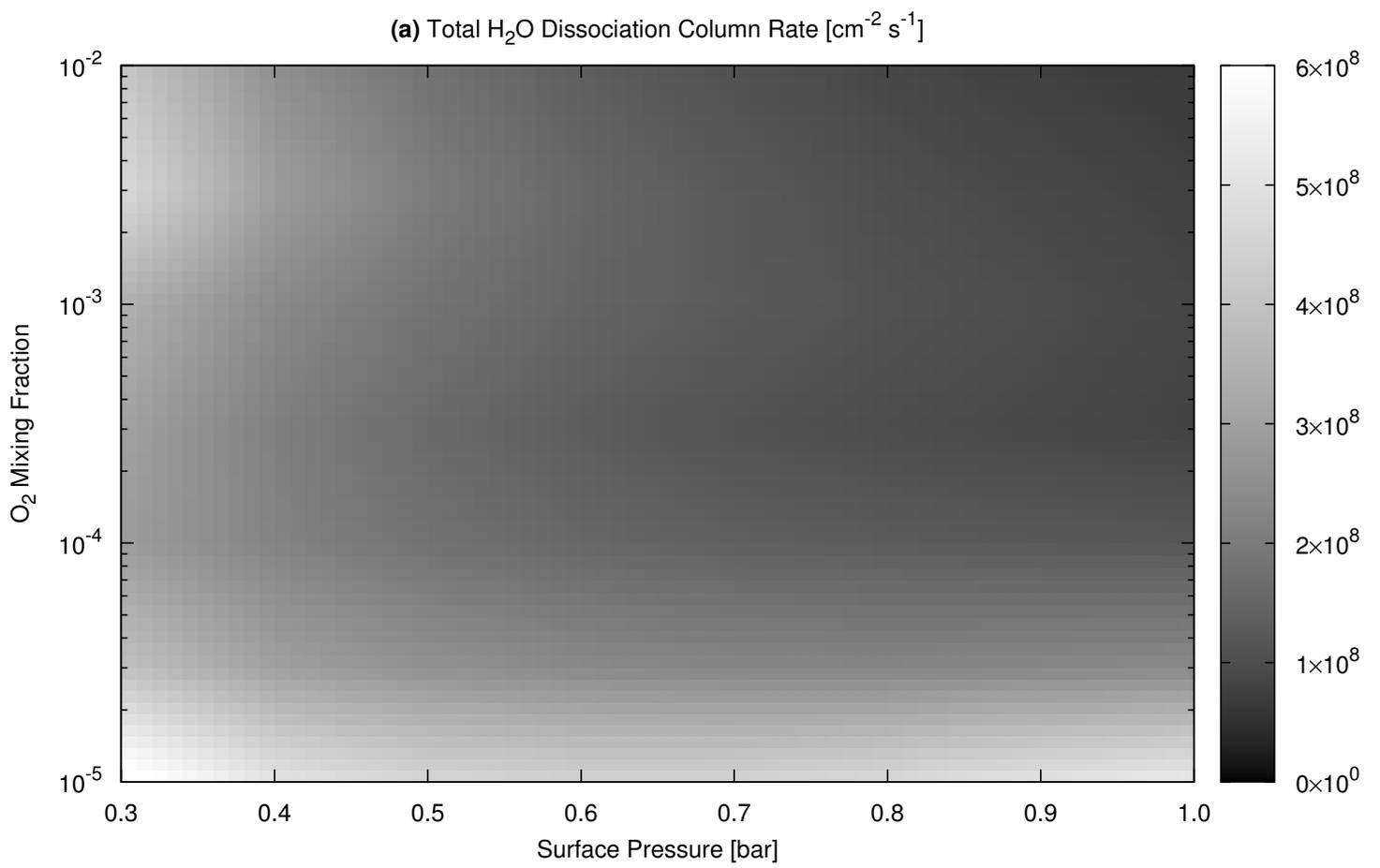

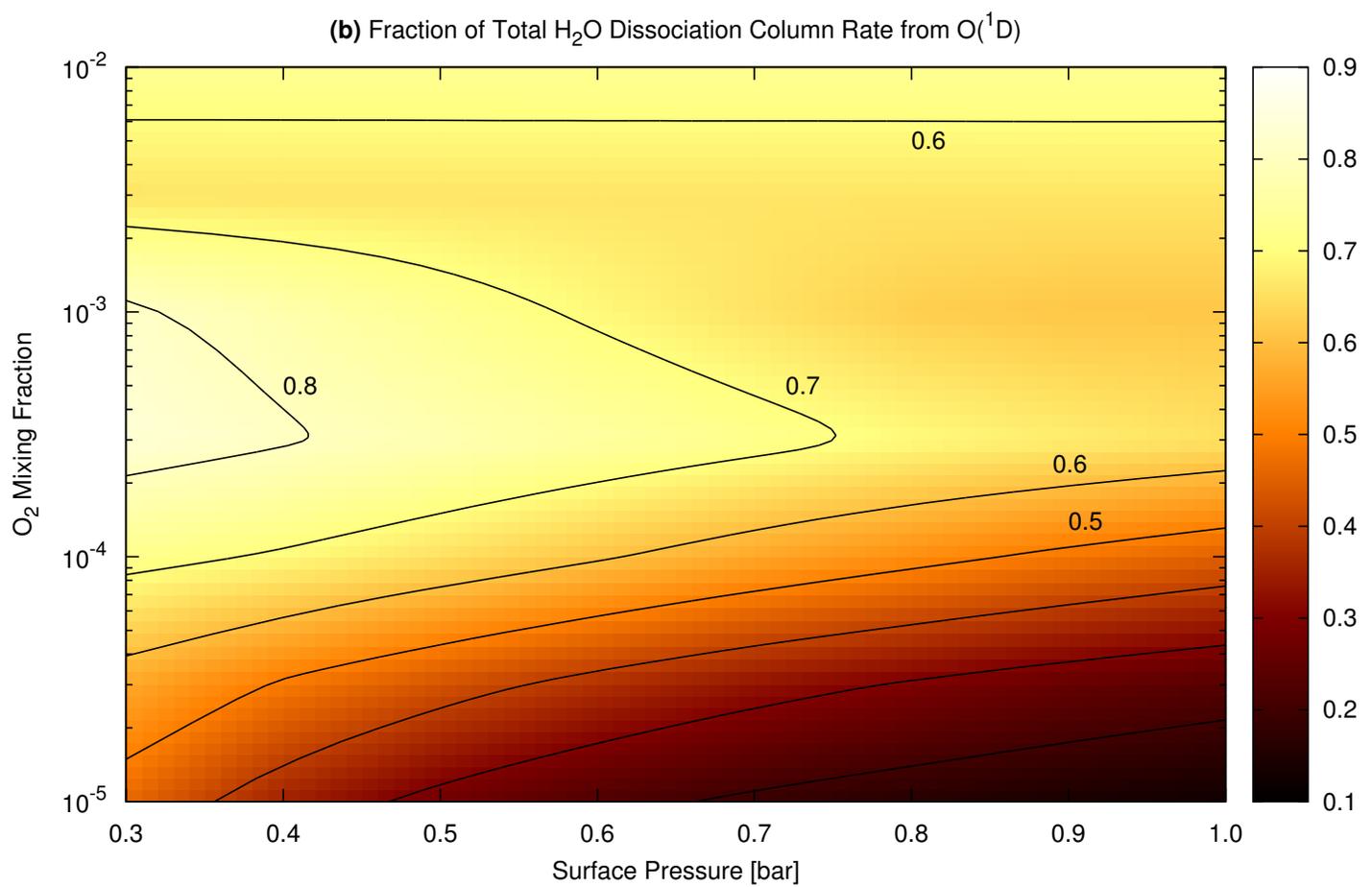

**(b)** Fraction of Total $H_2O$ Dissociation Column Rate from $O(^1D)$

Figure8b Grey

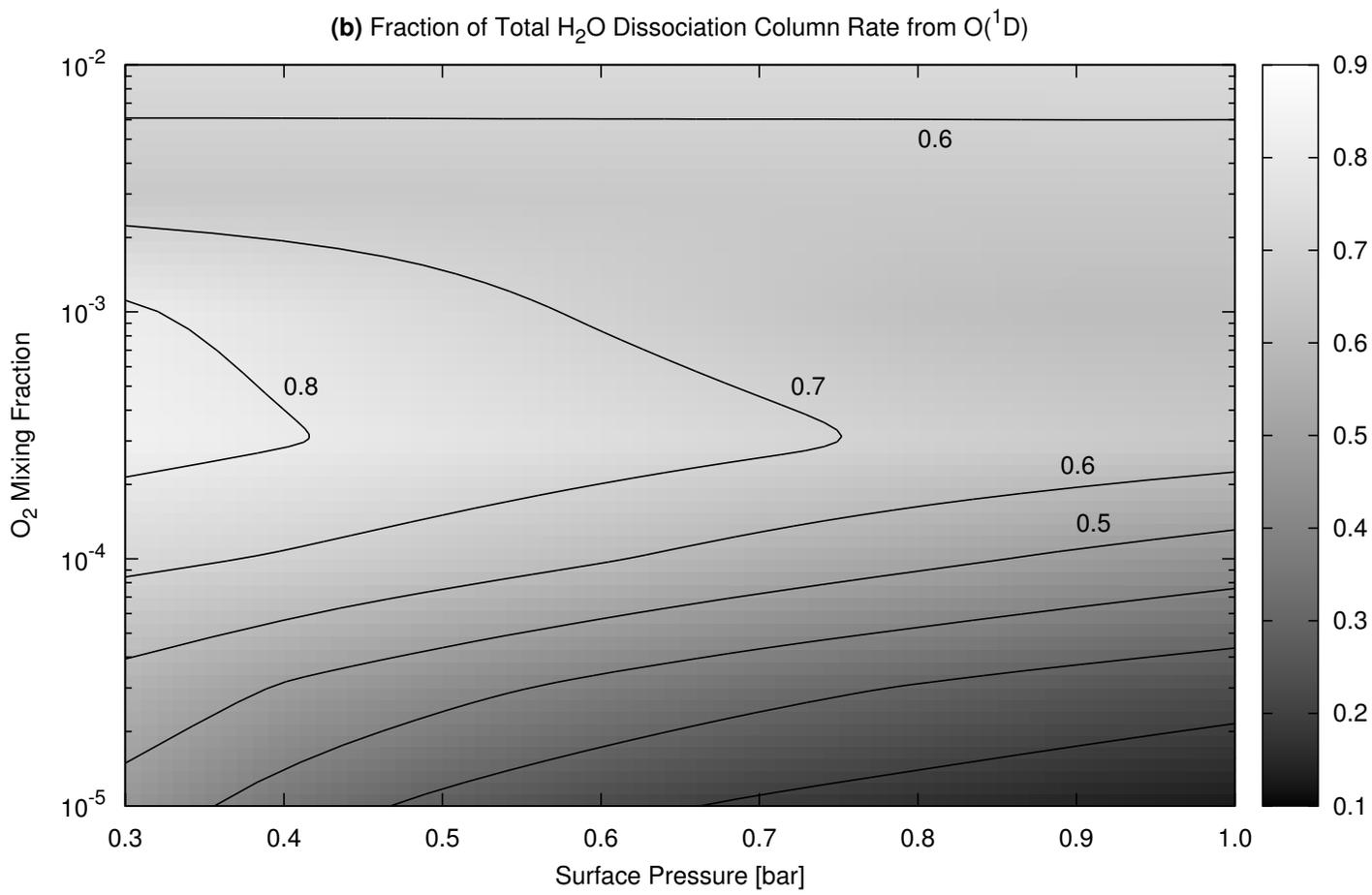

**(b)** Fraction of Total $H_2O$ Dissociation Column Rate from $O(^1D)$

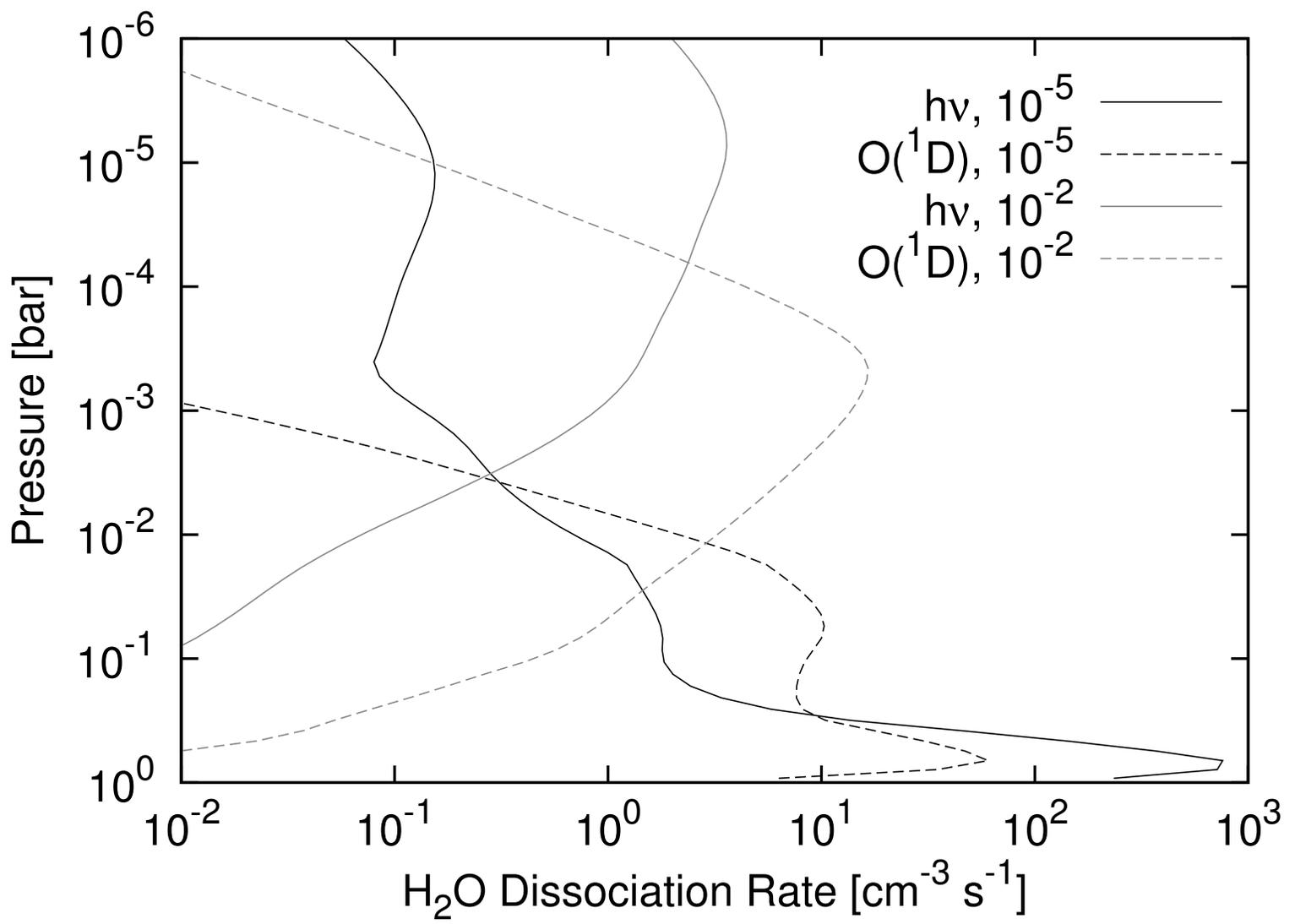